\documentclass[journal]{IEEEtran}
\usepackage{cite}
\ifCLASSINFOpdf
  \usepackage[pdftex]{graphicx}
\else
\fi
\usepackage{amsmath}

\usepackage{caption}
\usepackage{subcaption}
\usepackage{mathtools}

\DeclarePairedDelimiter\floor{\lfloor}{\rfloor}

\usepackage[ruled, linesnumbered, vlined]{algorithm2e}
\usepackage[noend]{algorithmic}
\usepackage{url}
\usepackage{balance}
\usepackage{color,soul}

\usepackage{multirow} 
\usepackage{pifont}

\usepackage[colorinlistoftodos,prependcaption]{todonotes}

\begin{document}

\title{Reliable and Distributed Network Monitoring via In-band Network Telemetry}

\author{
	\IEEEauthorblockN{Goksel Simsek\IEEEauthorrefmark{1}, Do\u{g}analp Ergen\c{c}, Ertan Onur\IEEEauthorrefmark{1}} \\
	\IEEEauthorblockA{\IEEEauthorrefmark{1}Department of Computer Engineering, \textit{Middle East Technical University}, TR} \\
	\IEEEauthorblockA{\IEEEauthorrefmark{2}Department of Computer Science, \textit{Universit{\"a}t Hamburg}, DE}\\
	\IEEEauthorblockA{\IEEEauthorrefmark{1}\{e2036598, eonur\}@ceng.metu.edu.tr}\\
	\IEEEauthorblockA{\IEEEauthorrefmark{2}\{ergenc\}@informatik.uni-hamburg.de}\\
}


\maketitle

\begin{abstract}

Traditional network monitoring solutions usually lack of scalability due to their centralized nature collecting heartbeats from all network components via a single controller.
As a solution, In-Band Network Telemetry (INT) framework has been recently proposed to collect network telemetry information more autonomously and distributedly by employing programmable switches.
However, it imposes further challenges to (i) find suitable INT paths to optimize the control overhead and information freshness and (ii) ensure reliable delivery of control information over multi-hop INT paths. 
In this work, we propose a monitoring scheme, reliable Graph Partitioned INT~(GPINT), by extending our previous work and integrating shared queue ring~(SQR) as a reliability feature against potential failures in network telemetry collection due to network congestion and link degradation that may cause loss of the visibility of the network. 
We implement our proposal in a recent data plane programming language P4, and compare it with traditional Simple Network Management Protocol~(SNMP) and also another state-of-the-art study employing Euler's method for INT path generation.
Our analysis first shows the importance of having a data recovery mechanism against packet losses under different network conditions. Then, our emulation results indicate that GPINT with reliability extension performs much better than its opponent in terms of telemetry collection latency and overhead monitoring scheme even under a high amount of packet losses.
\end{abstract}


\IEEEpeerreviewmaketitle

\section{Introduction} \label{sec:intro}

As a result of the integration of various new technologies and services, modern networks have become highly-complex ecosystems with a multitude of components. While data centers are enlarging to meet the need for exploding data-driven services, heterogeneity in the Internet of Things~(IoT) enforces the change of networking paradigms with new challenges~\cite{iot_survey}. %
Accordingly, the increasing complexity of networks requires efficient monitoring and management mechanisms to guarantee (i) reliable communication to detect and mitigate network failures and attacks, and (ii) to reconfigure the system in case of dynamically changing resource and traffic demands, especially for critical components and services.

In traditional networks, network devices usually require to be configured and managed manually via vendor-specific commands and policies.
The prominent network management protocols such as Simple Network Management Protocol (SNMP)~\cite{snmp}, NetFlow~\cite{netFlow} and sFlow~\cite{sflow_techreport} cannot offer an extensive solution for modern networks~\cite{traditional_monitoring_yang, traditional_monitoring_jammal}, such as 5G networks and cloud-based data centers~\cite{network_monitoring_sdn_review} since they require more autonomous and scaling solutions to deal with their increasing complexity~\cite{network_monitoring_sdn_review, traditional_scalability_pras}.
Software-defined Networking~(SDN) has been proposed to address some of the existing issues by introducing various link- and port-based telemetry collection mechanisms via a centralized controller~\cite{sdn_distributed, sdn_wireless_sensor}. 
However, polling telemetry measurements from each network node from a (logically) centralized controller does not scale well as it increases control overhead for measurement requests and responses ~\cite{sdn_monitoring_challenges, sdn_monitoring_sezer}. Besides, it imposes a strong dependency on (i) a centralized controller hindering more efficient use of data plane switches on monitoring and (ii) the features of SDN-specific protocols, e.g., OpenFlow~ \cite{openflow}, to collect only supported measures ~\cite{p4_review_kaur, programmable_review_bifulco, p4_review_exhaustive}.

\begin{figure}[!t]
	\centerline{\includegraphics[scale=0.565]{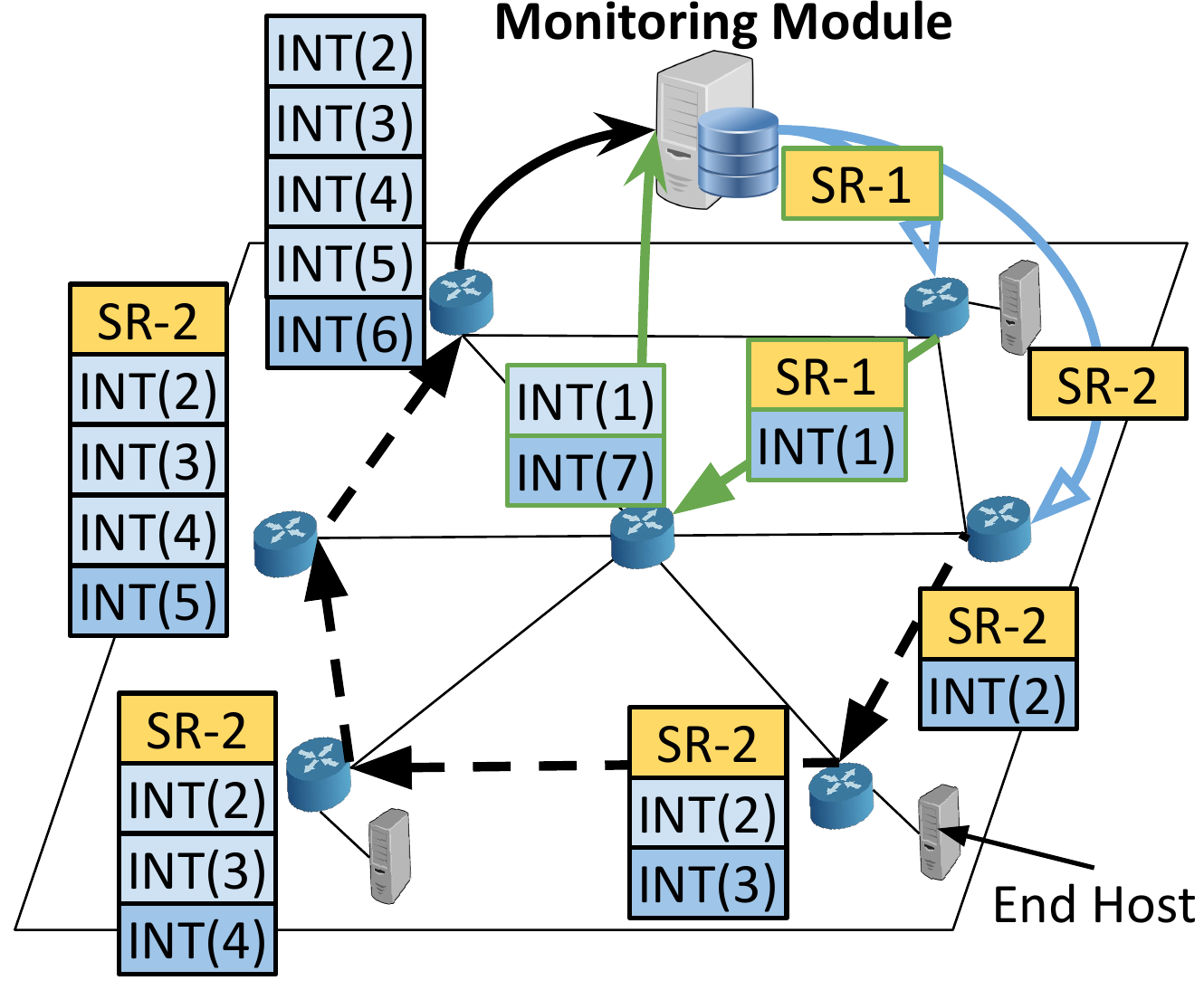}}
    \caption{An example usage of INT to cover every switch in the network.}
	\label{fig:scenario}
	\vspace{-6mm}
\end{figure}

To cope with such limitations, the P4 (Programming Protocol-independent Packet Processors) language~\cite{p4_paper} has emerged. It is a data plane description language that facilitates the customization of packet processing and forwarding pipelines to introduce new protocols or to customize packet processing on each data plane switch~ \cite{p4_review_exhaustive, programmable_review_kaljic}. Besides, P4 directly proposes a scalable telemetry mechanism, In-band Network Telemetry~(INT)~\cite{int_paper} to address the monitoring challenges introduced by polling-based methods, which are vastly employed in traditional network management protocols and also SDN. It enables the use of specific \textit{INT probes} to query telemetry information such as queuing latency and queue depth from P4-programmable switches. Once an INT probe is triggered and forwarded through a predetermined path, i.e., \textit{INT path}, each node through on the path extends this packet by appending its own measurements. 
Eventually, the destination node of an INT path receives the collective information of several nodes within a single INT packet without probing each node separately~\cite{INTPath}.

In Fig.~\ref{fig:scenario}, we depict an example use of INT telemetry collection session. In this example, a monitoring unit generates two probe packets, SR-1 and SR-2, that traverse through two disjoint INT paths covering all nodes to construct a global view. As SR-2 follows a longer path than SR-1, it reaches to the monitoring unit later and thus brings relatively older information. However, it also eases telemetry collection as a single INT packet collects extensive information with minimal overhead. Besides, all INT packets should be collected, even in case of link failures or packets losses, to maintain the current state of the network.

In our latest study, we proposed Graph Partitioned INT~(GPINT), a balanced INT path generation model~\cite{drcn_gpint}, to find the suitable INT paths minimizing the control overhead and the latency for telemetry collection. Here, we propose a reliability extension for our INT telemetry scheme that provides seamless recovery from link failures and potential INT probe losses by integrating shared queue ring (SQR)~\cite{sqr}. Our contributions are listed as follows.

\begin{itemize}
\item We develop a resilience mechanism for the INT monitoring scheme against packet loss and link failures employing SQR.
\item We design a complete network monitoring scheme including both control and data plane functionality and implement our proposal in P4. This includes all necessary monitoring modules and submodules as well as a protocol design with suitable packet structure and various temporal design parameters.
\item We evaluate our proposal under certain failure scenarios, comparing it to a pioneer INT-based telemetry collection study Pan et al.'s Euler method ~\cite{INTPath} and traditional SNMP in terms of efficiency and reliability.
\end{itemize}

The rest of the paper is organized as follows. Section~\ref{sec:probdef} gives the problem definition presenting the design requirements and the state of the art partially addressing them. Section~\ref{sec:system-design} presents the main control and data plane components of our monitoring framework, including the details of our previous work, GPINT~\cite{drcn_gpint}. In Section~\ref{sec:data-recovery}, we give the basics of SQR and explain our data recovery algorithm. Section~\ref{sec:results} presents our emulation environment and the experimental results. Lastly, Section~\ref{sec:conclusion} concludes the paper with potential extensions.

\section{Problem Definition} \label{sec:probdef}

In this section, we define our target problems listing the main requirements to design an INT-based monitoring scheme. Then, we review the state of the art with respect to the given requirements.

\subsection{Requirements} \label{sec:requirements}

There are three main requirements to ensure complete network visibility in an up-to-date and reliable manner. \\
\textbf{Requirement 1}. \emph{INT path(s) should traverse all nodes in the network}:
As the monitoring system should obtain a complete network view, generated INT packets should traverse all nodes in the network, i.e., each node should receive at least one INT packet to piggyback its measurements. To minimize the control overhead, there should be an optimum number of maximally-disjoint INT paths to cover the whole network. \\
\textbf{Requirement 2}. \emph{End-to-end latency of INT paths should be similar}: 
To have an up-to-date holistic view, all generated INT packets should be received concurrently, and it requires the use of \textit{balanced}, e.g., inducing similar latency or having similar lengths, INT paths. \\
\textbf{Requirement 3.} \emph{All INT packets should be delivered}: 
An unreliable control scheme can hinder the timely collection of telemetry in case of packet losses as even a single INT packet can carry multiple measurements depending on the length of its respective INT path. Therefore, INT monitoring should have required reliability mechanisms in place to maintain the current network view.

Consequently, although INT-based monitoring solves some issues of the traditional monitoring mechanisms, it brings the given challenges. Our previous work addresses the first and the second requirements ~\cite{drcn_gpint}. Accordingly, in this paper, our main goal is to extend that work to address the third requirement ensuring reliability in telemetry collection.

\subsection{State of the Art}

In traditional monitoring and management protocols like SNMP~\cite{snmp}, a controller requests statistics from several devices one by one in low frequencies~\cite{NetView}. Various approaches such as (i) sampling-based monitoring collecting measurements with a fixed sampling rate~\cite{netFlow, sflow, open_netmon, open_sample, payless, event_triggered_monitoring} and (ii) sketch-based monitoring offering compact data structure for summarized network measurements \cite{countSketch, countMinSketch, openSketch, univMon, elasticSketch, nitroSketch} suffer from high communication overhead and lack of collecting certain telemetry data. INT addresses those problems by embracing network programmability. INT-based approaches in the literature can be categorized into two: sampling-based and probe-based.

In sampling-based INT monitoring approaches, each switch is programmed to insert INT headers including network telemetry information to every received data packet. Before a packet is delivered to the destination host, the edge switch extracts the accumulated telemetry data and sends it to the network controller. In one of the earliest works, Kim et al.~\cite{int_paper} employ sampling-based INT to construct continuous latency plots specific to HTTP requests. Both IntMon~\cite{onos-based} and INTCollector~\cite{int_collector} design a data collection and analysis architecture as a service to improve the measurement analysis procedure. Hyun et al. \cite{realtime_intcollector}, propose a combination of two strategies, an INT management architecture built on top of INTCollector. Selective-INT~\cite{sINT} and Flexible Sampling-based INT~\cite{flex_int} show that adaptive INT insertion rate can achieve similar results compared to inserting INT to every packet but with much lower overhead. Differently, INT-Label~\cite{int_label} considers labeling state of switches with an adaptive rate.
Every switch maintains a labeling rate and accordingly inserts port statistics to the commercial traffic.
PINT~\cite{pint}, on the other hand, argues that inserting the complete device information at every hop induces high overhead and designs a probabilistic variation of INT that spreads out the information onto multiple packets. Its authors show that telemetry data approximation is sufficient for monitoring applications to perform similarly to collecting complete telemetry data.

The main problems with the sampling-based INT monitoring are (i) it couples data traffic with the control traffic, which mostly has different priorities and requirements in the network, and (ii) adds additional overhead to each data packet. Besides, only the switches actively forwarding packet sends their telemetry information at a certain with no latency guarantees. Therefore, the overall approach violates requirements 1 and 2. Besides, they do not offer any in-place reliability mechanism, which does not satisfy our third requirement.

Probe-based INT monitoring addresses the problems of sampling-based INT by introducing special INT packets to collect telemetry data from switches. It grants the flexibility of redirecting artificial packets and querying only the information mentioned in these probes in comparison to the previous approaches.
NetVision~\cite{NetVision} and INT-Path~\cite{INTPath}, are two of the prominent examples of probe-based monitoring. They first generate probe paths to direct INT probes via source-routing by leveraging Euler's and Hierholzer's algorithms~\cite{euler_theory}, which give a theoretical minimum number of non-overlapping paths covering the whole network. The performance of INT-Path relies on the number of odd degree vertices in the network, limiting the applicability and scalability. Therefore, it satisfies the first requirement but the second and third requirements are missing. Ariel et al.~\cite{near_optimal_int} improves this approach in terms of control overhead introduced by the telemetry collection probes. Bhamera et al.~\cite{int_opt} propose INTOpt by employing probe-based INT to monitor Service Function Chains (SFCs).
They propose a simulated annealing-based random greedy heuristic to address a range of requirements of different service functions.
Marques et al.~\cite{marques} consider multiple monitoring applications requesting link statistics, similar to INTOpt.
They utilize simple INT paths to cover every interface, i.e., links between switches, in the network which drastically increases the number of generated paths.
Besides, joint paths increase the overhead on switches, and in the case of high-frequency monitoring applications, the overhead may become significant. Accordingly, it satisfies the first requirement in an inefficient way while missing the second and third requirements.
NetView~\cite{NetView}, on the other hand, utilizes simpler paths to cover all switches instead of interfaces and collects additional telemetry data to estimate link utilization.

Our previous work, GPINT ~\cite{drcn_gpint}, is finding a $n$ nearly-equal sized INT paths covering the whole network with maximally-disjoint paths satisfying the first and the second of the given requirements. Here, we improve that for resilience against potential link failures and packet losses that may hinder the overall monitoring process and accordingly address the third requirement as well.

\section{Overview of the Monitoring Framework} \label{sec:system-design}

In this section, we present the design of our monitoring module, data plane implementation, and control packet structure exchanged between the monitoring module and data plane components.

\subsection{Monitoring Module: Control Plane} \label{sec:controller-design}

\begin{figure}[t]
	\centerline{\includegraphics[width=\columnwidth]{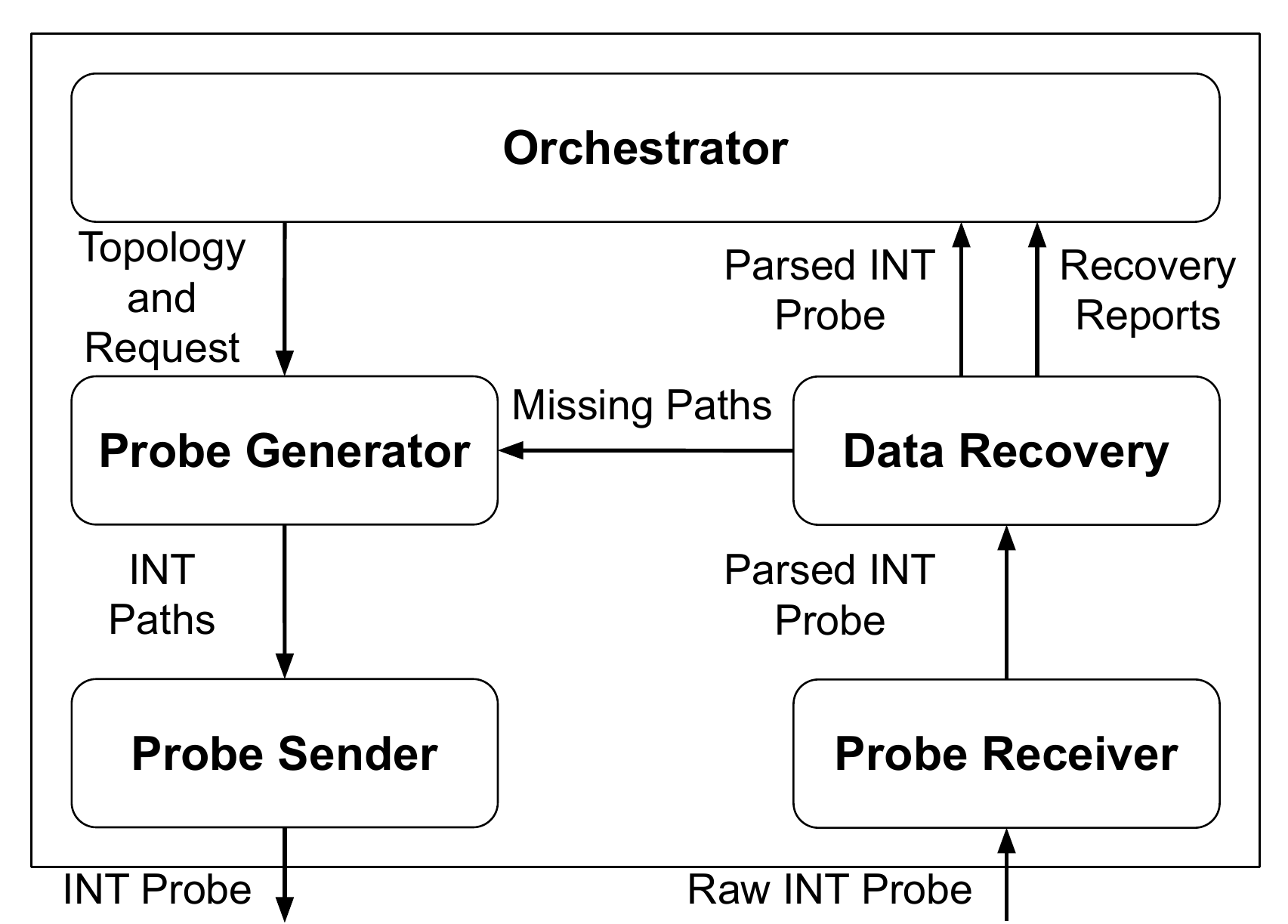}}
    \caption{Proposed monitoring module used in P4 emulations}
	\label{fig:inside-controller}
\end{figure}

Fig.~\ref{fig:inside-controller} shows our monitoring module architecture.
In the module, the \textit{orchestrator} initiates a monitoring request for certain telemetry information, e.g., rule-hit ratio or packet-per-port counter, from a network controller.
It then calls the \textit{probe generator} loading with overall network topology, e.g., an undirected graph, to generate INT paths. Our heuristic, GPINT, is implemented in this module and briefly explained at the end of this section.
Then, \textit{probe sender} constructs INT probe packets injecting decided INT paths to source-routing headers and transmits the packets. 

When an INT probe is received after traversing the network, it is first parsed by \textit{probe receiver}.
The \textbf{data recovery} module detects and mitigates a potential loss of telemetry data under certain conditions. It also alarms \textit{probe generator} for potentially faulty packets. The probe packet is lastly delivered to the \textit{orchestrator} to extract the actual telemetry data.

Note that the data recovery module has two complementary components in the monitoring module and on the data plane, i.e., on each switch. \\
\textbf{Graph Partitioned INT (GPINT)}: GPINT extends Kernighan-Lin graph partitioning algorithm~\cite{kernighan1970efficient} to generate multiple INT paths with similar lengths that eventually cover the whole network.
We utilized the (i) initial partitioning and (ii) exchange phases of the algorithm to satisfy the first and the second requirements given in the previous section.

\begin{figure}[h]
  \begin{subfigure}{0.45\columnwidth}
    \includegraphics[width=\textwidth]{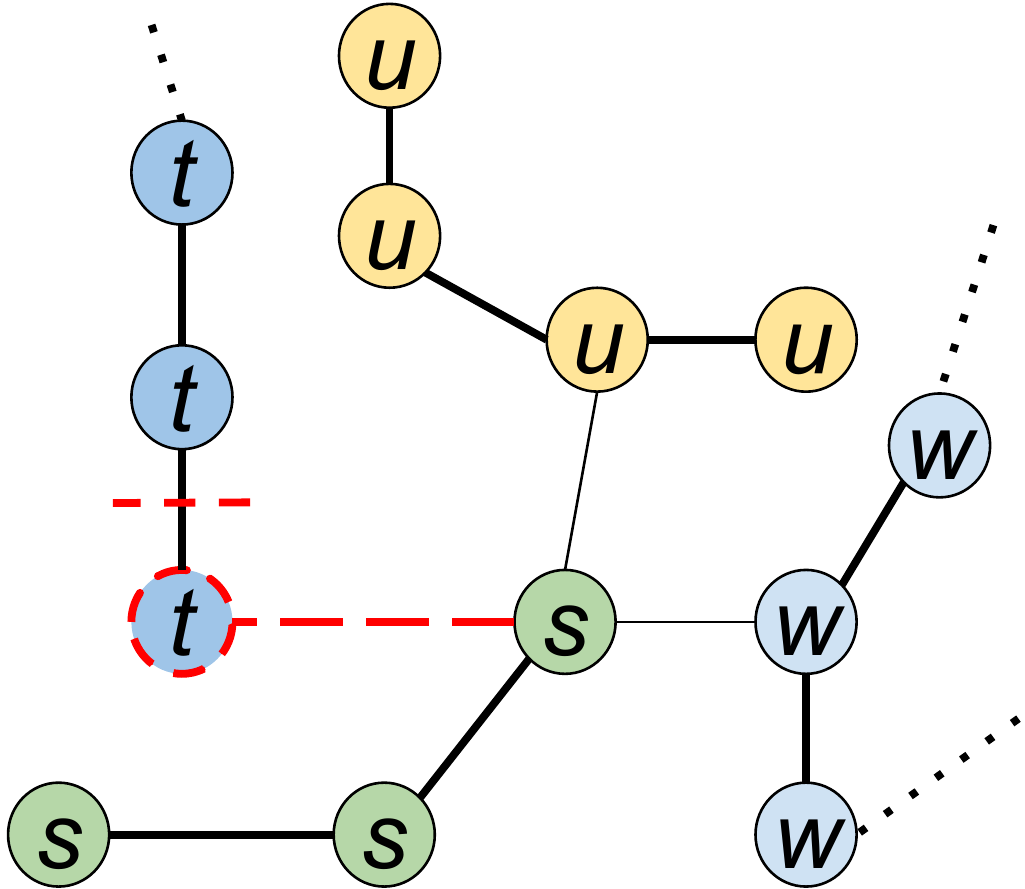}
    \caption{Option (ii-a)}
    \label{fig:example-exchange-2a}
  \end{subfigure}
  \begin{subfigure}{0.45\columnwidth}
    \includegraphics[width=\textwidth]{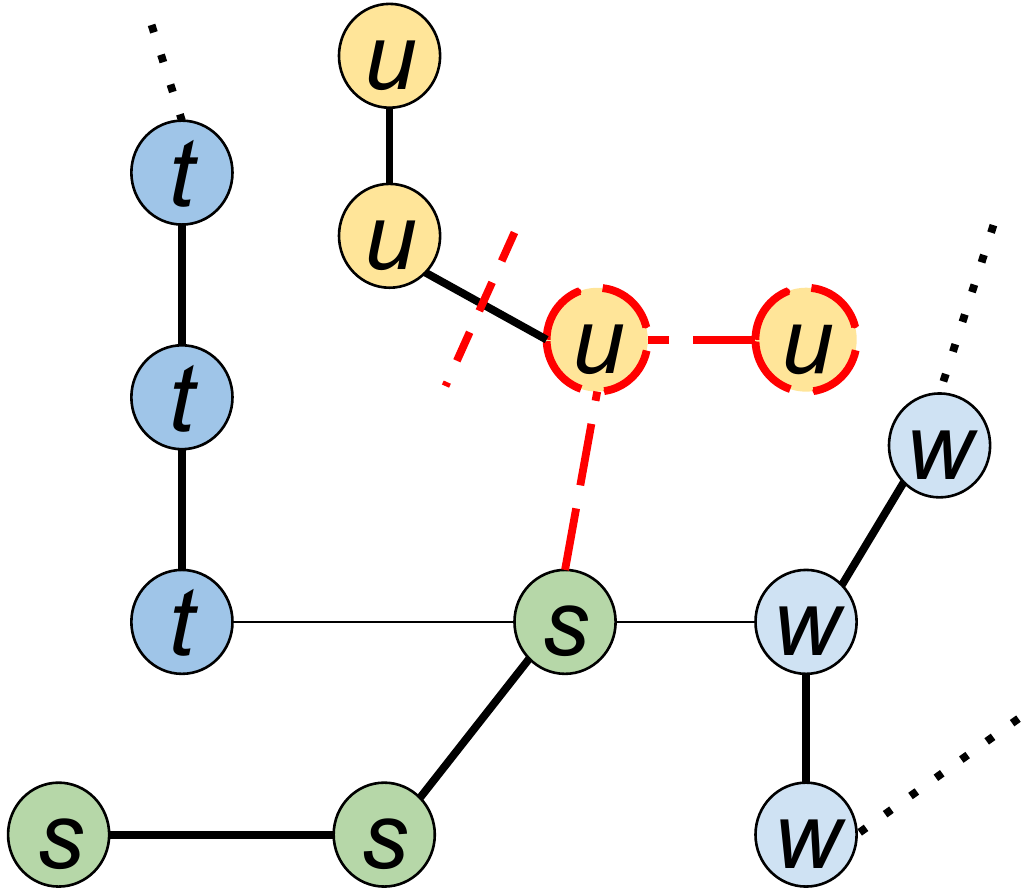}
    \caption{Option (ii-b)}
    \label{fig:example-exchange-2b}
  \end{subfigure}
\center{
  \begin{subfigure}{0.45\columnwidth}
    \includegraphics[width=\textwidth]{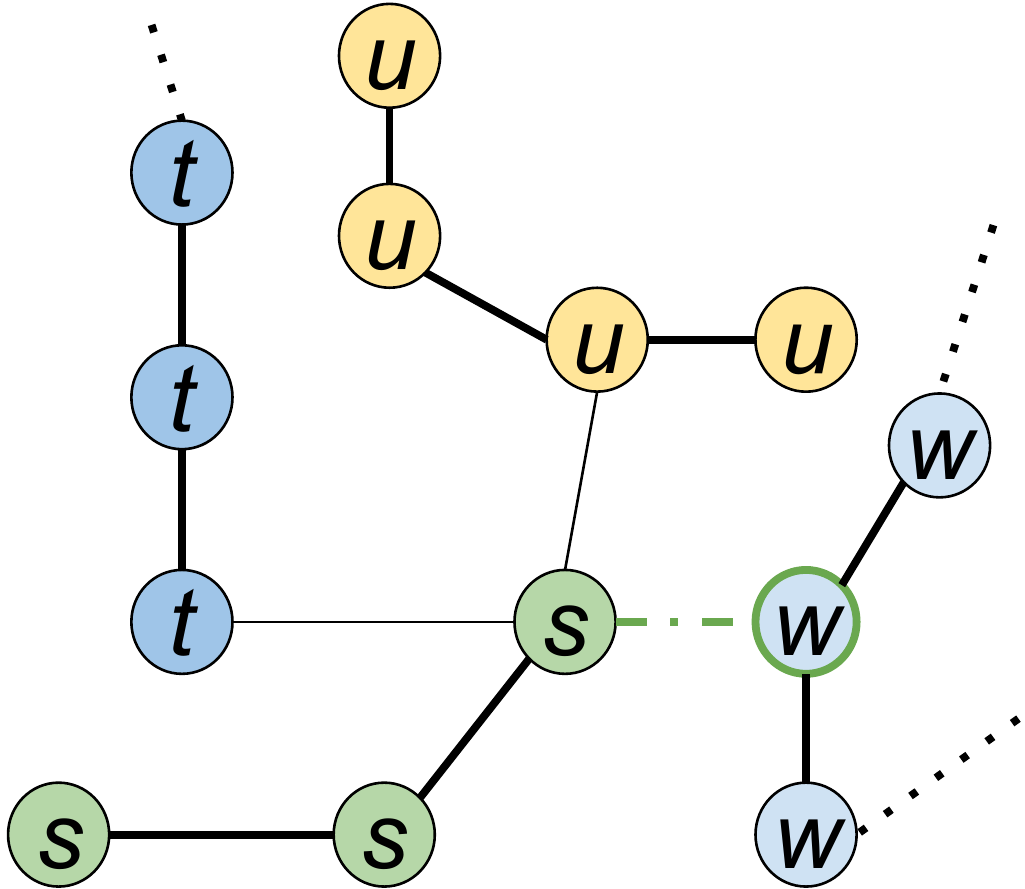}
    \caption{Option (ii-c)}
    \label{fig:example-exchange-2c}
  \end{subfigure}
}
  \caption{Illustration of available exchange options from path \textit{s}' point of view.}
  \label{fig:example-exchange}
\end{figure}
In the initial partitioning phase, for a given undirected graph $G(V, E)$ and the number of partitions to be generated $k$, the GPINT orders nodes by their degree and forms paths with approximately $\frac{|V|}{k}$ nodes each.
In the exchange phase, the GPINT transfers the nodes between neighbor paths, i.e., intersecting or having neighbor nodes, to satisfy the given requirements. Fig.~\ref{fig:example-exchange} shows different scenarios for such node transfers. In the scenarios, there are four different paths $s$, $t$, $u$, and $w$.
For the first method shown in Fig.~\ref{fig:example-exchange-2a}, GPINT tries to transfer one of the endpoints in $t$ to $s$, which can only happen if two nodes are connected and belong to different paths.
In Fig.~\ref{fig:example-exchange-2b}, it assigns a segment, i.e., two nodes from $u$ to $s$, instead of a single endpoint node.
In the last scenario, GPINT assigns a node to both paths $s$ and $w$ and constructs overlapping paths, which can be the case to form longer paths traversing the whole network at a cost of collecting the telemetry information of the same node twice.
The strategies in those scenarios are applied to every unvisited node in the graph measuring their impact on our objectives given in the requirements. This procedure continues until there is no improvement left for node transfers between $k$ constructed paths.
The formal definition of the algorithm and further complexity discussion can be found in~\cite{drcn_gpint}.

\subsection{INT Probe Packet Structure} \label{sec:packet-design}

\begin{figure}[t!]
	\centerline{\includegraphics[width=\columnwidth]{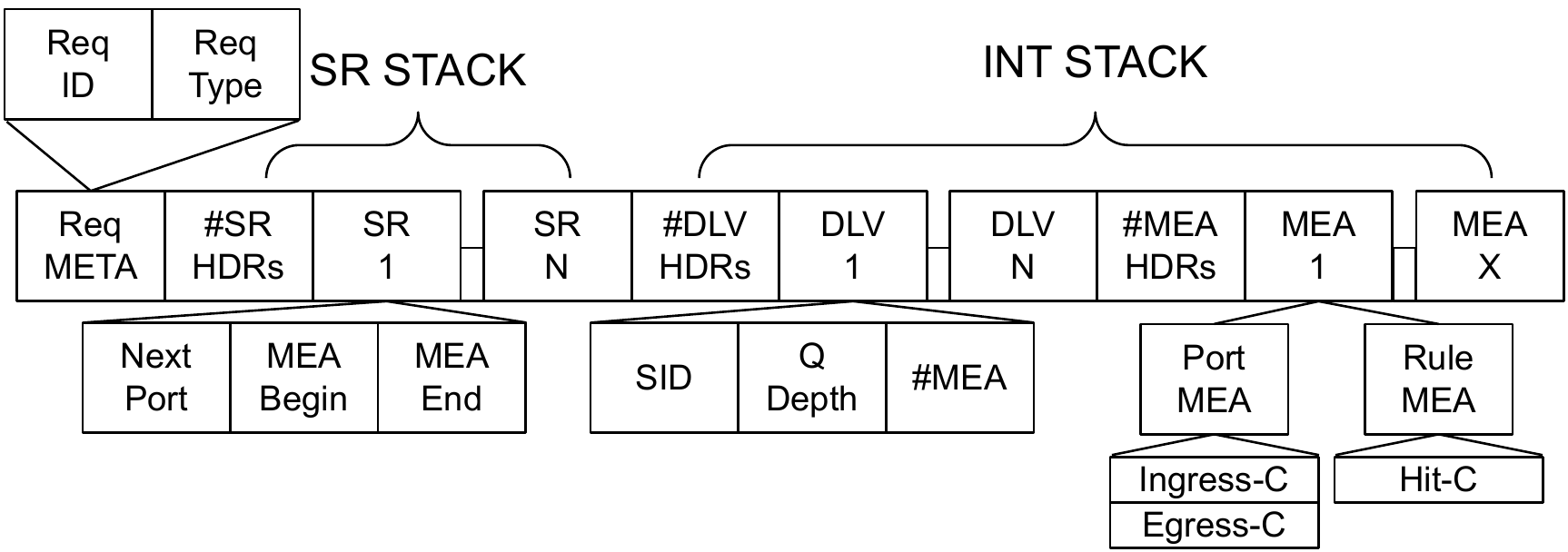}}
    \caption{INT probe packet layout}
	\label{fig:packet-layout}
\end{figure}

INT probe packets are exchanged between the monitoring module and edge switches, e.g., delivering the accumulated telemetry information to the monitoring module. Fig.~\ref{fig:packet-layout} shows the layout of those packets.

The header layout consists of three different partitions: \textit{request meta~(Req META)}, \textit{source-routing~(SR) stack}, and \textit{INT Stack}.
The \textbf{request meta} contains two fields: \textbf{request ID~(15 bits)} represents the unique identifier of an INT probe.
The \textbf{request type~(1 bit)} field specifies the requested telemetry information. 
We implement two types of configurable telemetry information to be carried by INT probe packets: rule-hit counters and port-based packet counters. They measure the number of hits for a certain forwarding rule on a switch and the number of packets received from each port of a switch, respectively.

The \textit{SR stack} accommodates the number of the SR headers~(16 bits, \#SR HDRs) and a list of SR headers~(SR 1-N), containing the next port information~(8 bits) and measurement (MEA) range~(16 bits). 
While the next port represents the next switch that the INT probe will be forwarded, the MEA range specifies the requested portion of telemetry information in the data plane registers of the respective switch. This enables the monitoring module to request partial information per switch flexibly.

Lastly, \textit{INT stack} includes device-level information~(DLV) and the measurements~(MEA). 
DLV further consists of the number of DLV headers~(16 bits, \#DLV HDRs) and a list of them. Each DLV header~(DLV 1-X) has a switch ID~(8 bits, SID), a queue depth~(8 bits, Q depth), and the number of measurements~(16 bits, \#MEA) collected from the switch with the respective ID according to MEA range.
Similarly, measurements contain the number of measurements~(16 bits) and the telemetry information of each switch on the INT path of the respective probe.

\subsection{Packet Processing Pipeline: Data Plane} \label{sec:dataplane-design}

In this section, we explain our data plane implementation that processes INT probe packets.

\begin{figure}[h]
	\centerline{\includegraphics[width=\columnwidth]{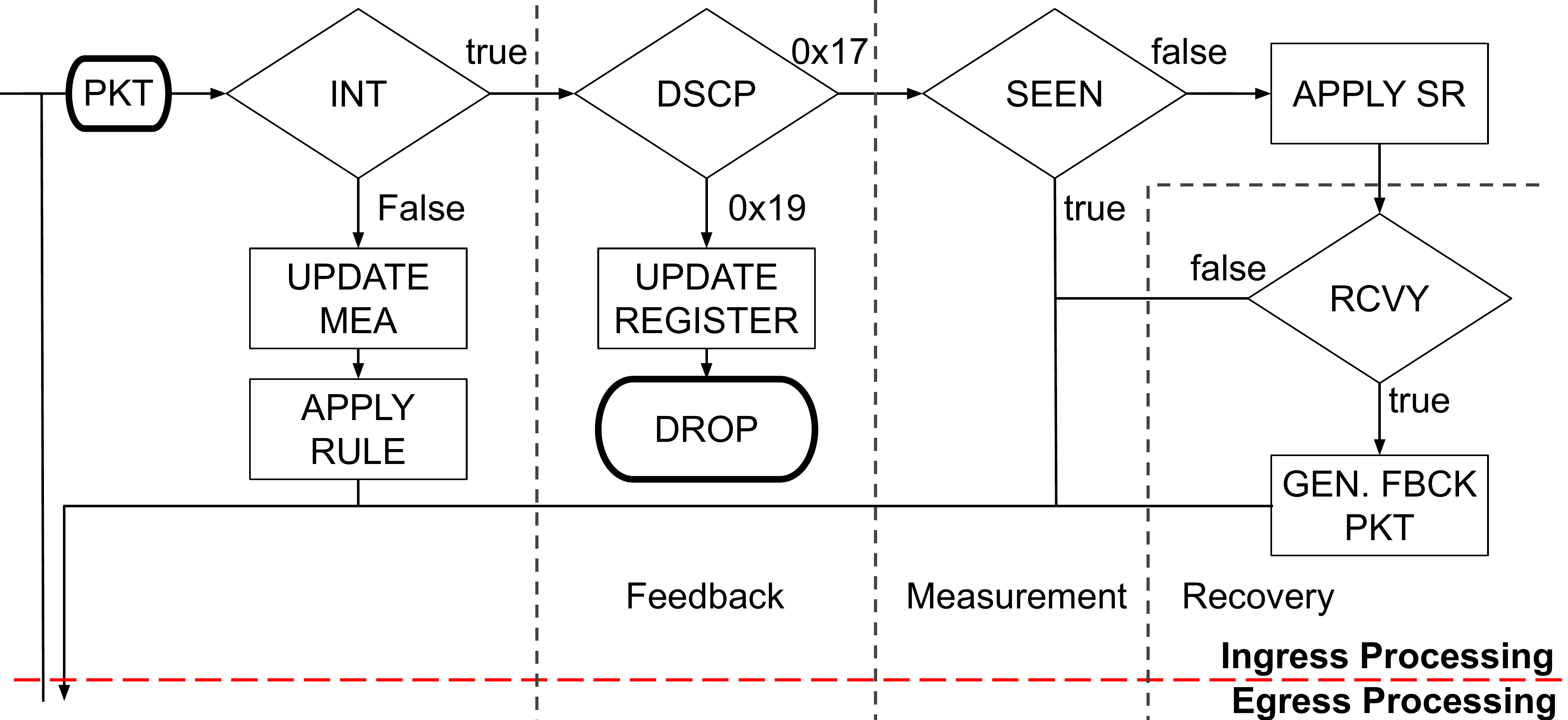}}
    \caption{The flowchart of the ingress pipeline}
	\label{fig:ingress-pipeline}
\end{figure}

When it is received, a packet is first distinguished as an INT probe by the Differentiated Services Field Codepoints~(DSCP) field of IPv4 header: $0x17$ for the probe with measurements and $0x19$ recovery feedback. INT probes are parsed into the given stacks and forwarded to the ingress pipeline, which is shown in Fig.~\ref{fig:ingress-pipeline}. For non-INT probes, the switch updates its respective packet counters and sends the packet to the egress pipeline. 

\begin{figure*}[t]
	\centerline{\includegraphics[width=\textwidth]{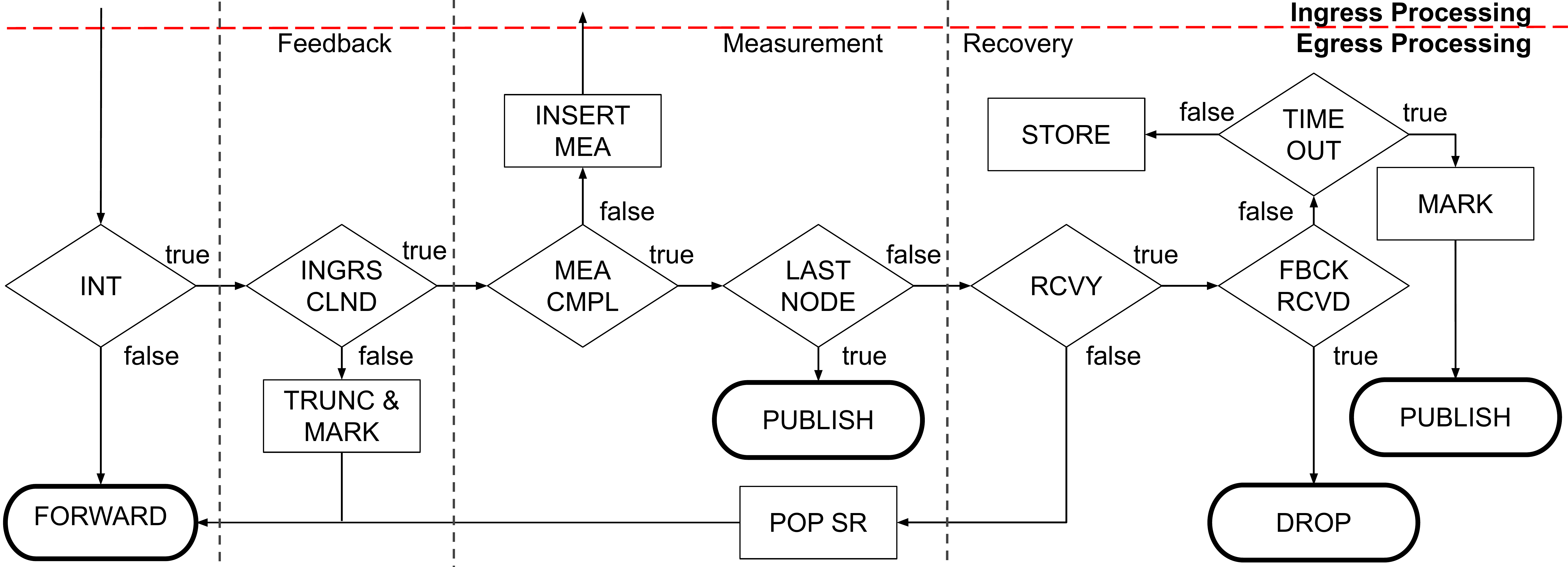}}
    \caption{The flowchart of the egress pipeline}
	\label{fig:egress-pipeline}
\end{figure*}

The probes with measurements, i.e., DSCP$=0x17$, are circulated between ingress and egress pipelines within the custom measurement ranges given in the respective SR header for the switch to parse the whole packet and insert requested metrics.
If the same packet is received before, e.g., because of a loop in the network, or the recovery mode is disabled on the switch, it is directly passed to the egress pipeline. Otherwise, firstly a feedback packet is generated to be used within the recovery module, whose mechanism is explained in more detail in the next subsection. Then, both the probe packet and the feedback packet are passed to the egress pipeline.

Fig.~\ref{fig:egress-pipeline} summarizes the egress pipeline.
Non-INT packets are directly forwarded to their destination ports.
The feedback packets generated by the ingress pipeline are truncated from unnecessary header parts and the DSCP field is tagged with~$0x19$.
For the measurement probes, the pipeline inserts the respective measurement for the requested counter if the MEA range limit has not been reached.
Then, according to the headers in \textit{SR stack}, the probe packet is forwarded whether (i) directly to the monitoring module if there are no more SR headers or (ii) to the next switch determined by the SR header, which is removed before the probe is being forwarded.
Note that a clone of this packet is preserved for a fixed timeout at this stage until a feedback packet is received by the next hop if the recovery mode is active. If it is not received at all, the packet's DSCP field is marked with $0x21$, and sent to the monitoring module alerting a potential network failure. Otherwise, the stored packet is dropped.

Note that we also implement an artificial loop between ingress and egress pipelines to process the probes with non-fixed measurement ranges and a list of headers since there is no built-in looping mechanism within the pipelines. For a better understandability, we have not discussed the implementation details further in this part.

\section{Reliable In-band Network Telemetry Collection} \label{sec:data-recovery}

Loss of even a single INT probe packet due to node or link failures can result in a lack of visibility for a large portion of the network as it carries accumulated information rather than a heartbeat for a single component. In this section, we present our packet recovery mechanism integrated into the INT monitoring scheme employing shared queue recovery~(SQR)~\cite{sqr} in the data plane for programmable networks.

\subsection{Overview of SQR}
SQR masks the failure in place, e.g., on data plane switches, so that the data traffic can be forwarded seamlessly without alarming the points or reinitiating packet transmission from the beginning. 
Eventually, it provides better link utilization and QoS together with increased reliability of the communication.

The main logic of SQR is as follows. Each programmable switch store a copy of every data packet before forwarding them to the destination address. Once a packet is received by the next hop switch, it generates a feedback packet in the reverse direction, e.g., to the immediate sender of the respective packet, to acknowledge the reception. When the feedback packet arrives, the switch discards the data packet being stored. If it is not within a fixed timeout, the respective switch alerts the network controller or the monitoring systems for a potential link or node failure occurring on the next hop switch. Moreover, that switch keeps saving the following data packets that would be normally forwarded through the problematic link/node until a controller or network administrator configures a new forwarding rule. Then, the traffic is rerouted according to the updated rule.

\subsection{Data Recovery Module}
In our INT monitoring scheme, we employ SQR such that a switch keeps a copy of received INT probe packets before forwarding them until the next hop switch (e.g., according to the SR headers) sends a feedback message. If it is received within a timeout period, the copy is discarded. Otherwise, the sender switch deduces a potential failure on the data plane and sends stored probe packets directly to the monitoring module of the system. This module then initiates rerouting by updating the forwarding rules in the respective switch.

\begin{figure}[h]
	\centerline{\includegraphics[width=\columnwidth]{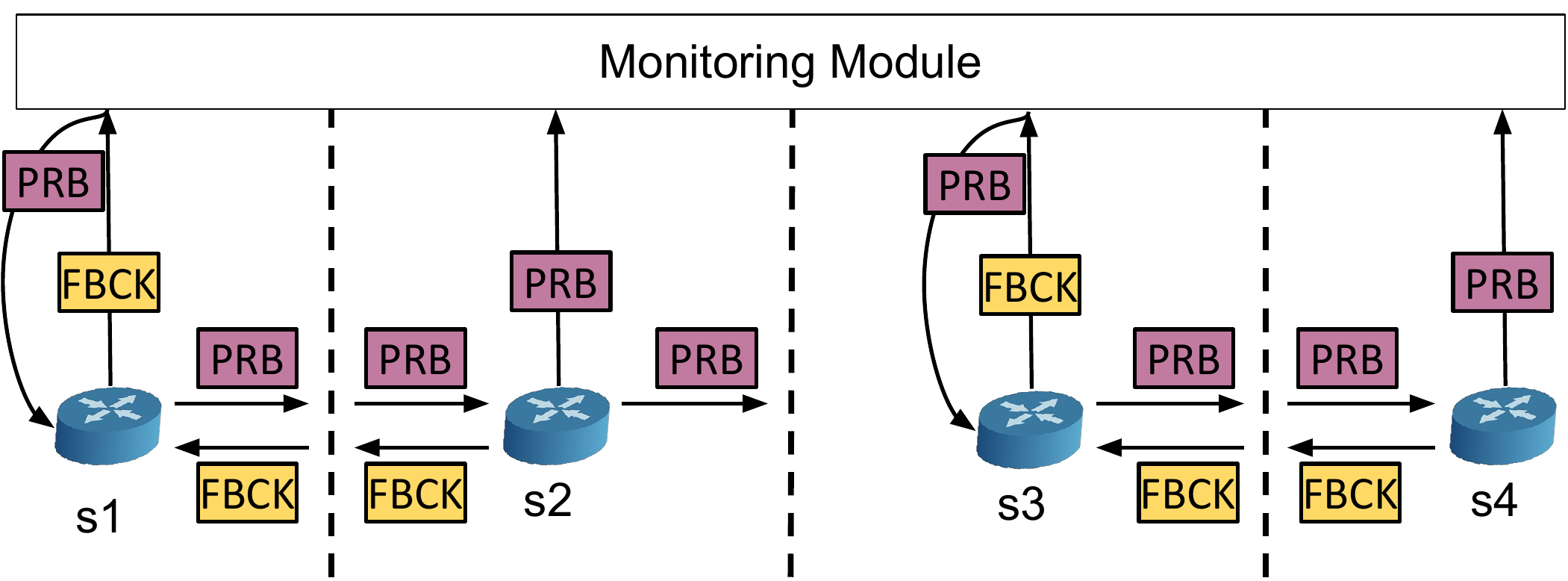}}
    \caption{An example scenario for data recovery}
	\label{fig:data-recovery-overview}
\end{figure}

Fig.~\ref{fig:data-recovery-overview} shows a data recovery scenario. In the figure, the events happening in each switch, e.g., receiving and sending INT probe~(PRB) and feedback~(FBCK) messages, are divided by dotted lines.
Here, the monitoring module initiates telemetry collecting by sending an INT probe packet to the edge switch $s_1$ (assuming that $s_1$ is the first switch of an INT  path) keeping a copy of the probe for $t_f$ seconds, which is the feedback timeout. 
If no feedback message is yet received after a number of recovery attempts, $r_a$, to initiate probing on $s_1$, the monitoring module marks the switch as unresponsive and uses another INT path instead, e.g., via $s_2$.
However, in this case, $s_1$ sends back a feedback probe containing only \textit{request meta} and the monitoring module deduces that $s_1$ is responsive and marks it so.
When $s_1$ receives the probe packet, it (i) inserts its measurements to the respective field of the packet, (ii) stores a copy of this updated probe, (iii) and sends it to the next hop in source-routing headers. It then waits for $t_f$ seconds to get a feedback message before discarding the copy.
$s_2$ sends the feedback and repeats the same process. However, in the figure, $s_3$ cannot send the feedback within $t_f$ seconds, and thus $s_2$ forwards the INT probe with partial telemetry information, e.g., from only $s_1$ and $s_2$, directly to the monitoring module. 
This module deduces a potentially broken INT path as the probe is not sent by the last hop in SR headers, e.g., $s_4$ in this example, but by an intermediate node. 
It then tries to initiate another INT probing from $s_3$, marking the connection between $s_2$ and $s_3$ as unavailable. However, it is also possible that the feedback of $s_3$ is only delayed because of, for instance, congestion on that link. If the first probe packet still completes the INT path and is delivered by $s_4$, the monitoring module updates the connectivity status between $s_2$ and $s_3$, e.g., not unavailable but congested.

Note that it can be easily costly to store all the incoming packets, especially when the timeout is not sufficiently small. We employ SQR for only INT packets to ensure the reliability of our monitoring mechanism. However, detecting a potential failure during the monitoring process also leads to the overall recovery process of the respective components that would prevent the loss of actual data packets as well.

In the rest of this subsection, we explain our implementation details algorithmically, discuss the impacts of different design parameters, and the potential limitations.

\subsection{Implementation Details} \label{sec:recovery-implementation-details}

In this section, we present the implementation details of our recovery approach. Algorithm~\ref{alg:recovery-main-loop} shows a pseudo-code for the main loop of the data recovery submodule of our monitoring module.
The data recovery module is configured with the parameters, $P_{INT}$, $F$, $Q$, and $\mathcal{A}$. 
$P_{INT}$ contains the information of the probes for each candidate INT path, which include (i) a unique identifier, (ii) time to initiate probing set by the \textit{orchestrator}, (iii) most recently observed path index, (iv) complete INT path, i.e., a source-route, and (v) an expected deadline to be received back with the telemetry measurements. 
$F$ represents a list of pending feedbacks containing the identifiers of each INT path to keep track of if a probe is received completing the whole path.
$Q$ is a priority queue where the INT probes are stored by their initiation time.
Lastly, $\mathcal{A}$ is a generic feedback list to record the deduced component status, e.g., a congested link or failed node.

\begin{algorithm}[t]
 \caption{Data Recovery Loop}
 \label{alg:recovery-main-loop}
 \algsetup{linenosize=\small}
 \small
 \KwData{$P_{INT}$, $F$, $Q$, $\mathcal{A}$}
 \KwResult{$\mathcal{A}$}
  \While{probe received or timeout raised}{
	\eIf{timeout raised}{  
			\For {each $p_{id}^* \in F$}{
				$p$ $\leftarrow P_{INT}$.get($p_{id}^*$)\;
				\If{time.now() - $p$.init $> t_f$}{
					\If{$p \notin Q$} {
						$Q$.insert($p$)\;
						\If{$p$.attempts $ > r_a$}{
							$\mathcal{A}$.insert\_unavailable($p^{*}$)\;
							$p$.increment\_index()\;
						}
						$p$.increase\_attempt()	\;
					}
				}
			}
		process\_recovery\_queue($F$, $Q$)\;
  	}(\tcp*[h]{probe $p^*$ is received}){ 
  		$F$.remove($p_{id}^*$)\;
  		$p$ $\leftarrow P_{INT}$.get($p_{id}^*$)\;
		\uIf{$p_{dscp}^*$ is $0x17$}{
		  	$p$.finish()\;
  		}
  		\uElseIf{$p_{dscp}^*$ is $0x19$}{
  			\If{$p \notin Q$}{
  				$Q$.insert($p$)\;
  			}
  		 }
  		\Else(\tcp*[h]{$p_{dscp}^*$ is $0x21$}) {  
  			$p$.update\_index($p^*$)\;
  			$\mathcal{A}$.insert\_unavailable($p^*$)\;
  			\If {$p \notin Q$} {
  				$Q$.insert($p$)\;
  			}
  		 }
  	 }
 }
\end{algorithm}

The data recovery module periodically checks whether a timeout event is (asynchronously) raised or a measurement probe is received (line 1). 
When it does not receive a measurement probe for $t_{ro}$ seconds, i.e., the resolution time, the timeout sequence begins (line~2-12). Here, the module traverses over the pending feedbacks for the probes of all INT paths that do not return to the monitoring module with the measurements. For each pending probe feedback $p^*$, it first checks
if $t_f$ seconds passed, i.e., a feedback timeout, after it is being sent, which indicates timeout for this particular probe (line~5). If the probe is already in the $Q$, which indicates that the respective INT path for this probe is already marked incomplete; then it is added to the queue (line~7). The module then checks if the probe attempt limit $r_a$ is exceeded which leads to marking that individual switch unavailable and updating $\mathcal{A}$ (line~9). Afterward, the tracking index is incremented, which clears the $r_a$ and allows us to resume from the next hop in the path (line~10).
At the end of a timeout event, the whole recovery queue is processed via Algorithm~\ref{alg:recovery-processing-recovery-queue}.
 
If a measurement probe packet $p^*$ is received within the resolution time, firstly $F$ is updated by removing the respective $p_{id}^*$ (line~14). 
If the DSCP field of the packet is $0x17$, it means that the probe packet successfully traversed its path and we mark INT probe $p$ as finished (line~17). If the DSCP field is $0x19$, it indicates a feedback packet from the first switch on the path. The probe $p$ is then added to the recovery queue to wait for the measurements (line~20).

If the DSCP field is $0x21$, it is a notification packet to initiate recovery. In this case, the data recovery module identifies the starting index of $p$ that indicates the switch on the INT path that the probe is sent from. That is, for $p_{index} = i$, the monitoring module has previously sent this probe to the $i$th switch on its source route, i.e., decided INT path. According to the switch that has sent this recovery probe, the data recovery module updates $p_{index}$ to be sent through the next available switch on the INT path of the probe (line~22) and records the unresponsive switch/link to the $\mathcal{A}$ deducing from the INT path (line~23). The probe is then added to the recovery queue with an updated initiation time.

When all INT probes are collected, the loop is terminated for a telemetry collection session and the unavailable switches/links are sent to the \textit{orchestrator} for initiating an implemented recovery process. Then, the loop restarts periodically.

\begin{algorithm}[t]
 \caption{Process Recovery Queue}
 \label{alg:recovery-processing-recovery-queue}
 \algsetup{linenosize=\small}
 \small
 \KwData{$F$, $Q$}
 \KwResult{$F$ }
 \For {each $p \in Q$} {
	\If {time.now() $<$ p.deadline} {
		continue\;	
	}
	
 	$Q$.pop($p$)\;
 	
 	\If {not $p$.is\_finished()} {
 		$F$.insert($p$)\;
 		send\_probe($p$)\;
 	}
 }
\end{algorithm}

Lastly, Algorithm~\ref{alg:recovery-processing-recovery-queue} shows the process of the recovery queue that is utilized within the timeout sequence in Algorithm~~\ref{alg:recovery-main-loop} to restart the telemetry collection for the unresponsive probes. If the deadline for the probe $p$, e.g., the expected time to receive the measurements, has not come yet, the probe is skipped (line~2-3). Otherwise, if there is no measurement probe is received (i.e., not marked as finished), the probe is added to the $F$ and sent through the switch specified in its index (lines~5-7).

Note that the design parameters $t_f$ and $r_a$ are mostly determined by the characteristics of the network, e.g., link data rate, reliability, and have an impact on the control overhead and bandwidth utilization. For the evaluation, we have selected reasonable values for the considered network topologies, i.e., $t_f=1.5$~seconds and $r_a=10$. Additionally, the data recovery module is designed to be a plugin module for any orchestrator, considering the overhead it may bring. While the implementation is optimized to reduce its overhead (i.e., index tracking and periodic timeout checks), it may cause small interference with other tenant applications reside in the orchestrator. If such a case arises, the network maintainers can decouple the data recovery module and use it as a separate application.

\section{Results and Discussion} \label{sec:results}

In this section, we present our experimental results comparing GPINT with/without the data recovery extension with Pan et al.'s Euler method~\cite{INTPath}, which is another recent INT path-based monitoring scheme and traditional Simple Network Management Protocol~(SNMP)~\cite{snmp}. We evaluate latency, reliability, and control overhead for increasing data traffic, network size, and the probability of failure.

\subsection{Emulation Setup and Methodology} \label{sec:emulation-setup}

We implemented our monitoring framework using ETH Zurich's P4 repository~\cite{p4Learning} as it introduces various improvements to the original P4 simulation environment.
We used Mininet~\cite{mininet} as the emulation environment on a server with 6 cores 4GHz CPU and 32GB RAM. The orchestrator in this work can deploy at most 12 concurrent INT probes.

For network topologies, we used Watts-Strogatz random graphs translated to the networks with programmable switches. All switches and links are assumed to be identical.
The maximum INT probe packet size is set to 2200 bytes and the maximum queue size for a switch is 1000 packets to ensure that the amount of data traffic that we used in the experiments can cause congestion and eventually packet drops. Additionally, bounding the queue size helps us to limit the memory consumption of a single switch, so that we can increase the network size while still successfully conducting experiments. Note that those values can be adjusted according to the network characteristics. For instance, the maximum size of a probe packet depends on the longest path generated by the probe generators. In this work, this value is set to 195 to accommodate Euler's probe generator. Additionally, limiting the queue size helps to force the data plane to drop packets to represent a congested network. For all experiments, we deployed 60 hosts machines and emulated UDP traffic between randomly selected hosts with different data rates given in the respective evaluation scenarios.
Together with the data traffic, the monitoring module initiates INT probes and performs the same measurement loop 10 times.
To evaluate the performance of our data recovery scheme, we introduce link degradation, where each link (excluding orchestrator to switch links) has a probability of 5-20\% to drop any given packet in different scenarios. It is explained further in the related evaluation scenario.

All experiments are repeated 20 times and the average results are given with a 95\% confidence interval.

\subsection{Numerical Evaluation}

We first evaluate the impact of congestion on the packet drops to show the overall network behavior. Then, with and without our data recovery mechanism, we measure the latency of telemetry collection, the ratio of delivered data traffic and INT probes, and the successfully monitored network proportion under link degradation for increasing network size. Our overall goal is to evaluate the efficiency and reliability of the data recovery module.

We use different configurations of GPINT in terms of the number of INT paths $k$ s.t. GPINT-2, GPINT-3, and GPINT-5. $k$ is selected proportional to the network size s.t. $k=\floor{\frac{2|V|}{10}}$, $k=\floor{\frac{3|V|}{10}}$, and $k=\floor{\frac{5|V|}{10}}$, respectively, where $|V|$ is the number of switches.

\subsubsection{The Impact of Congestion}

\begin{figure}[t]
  \begin{subfigure}{0.49\columnwidth}
    \includegraphics[width=\textwidth]{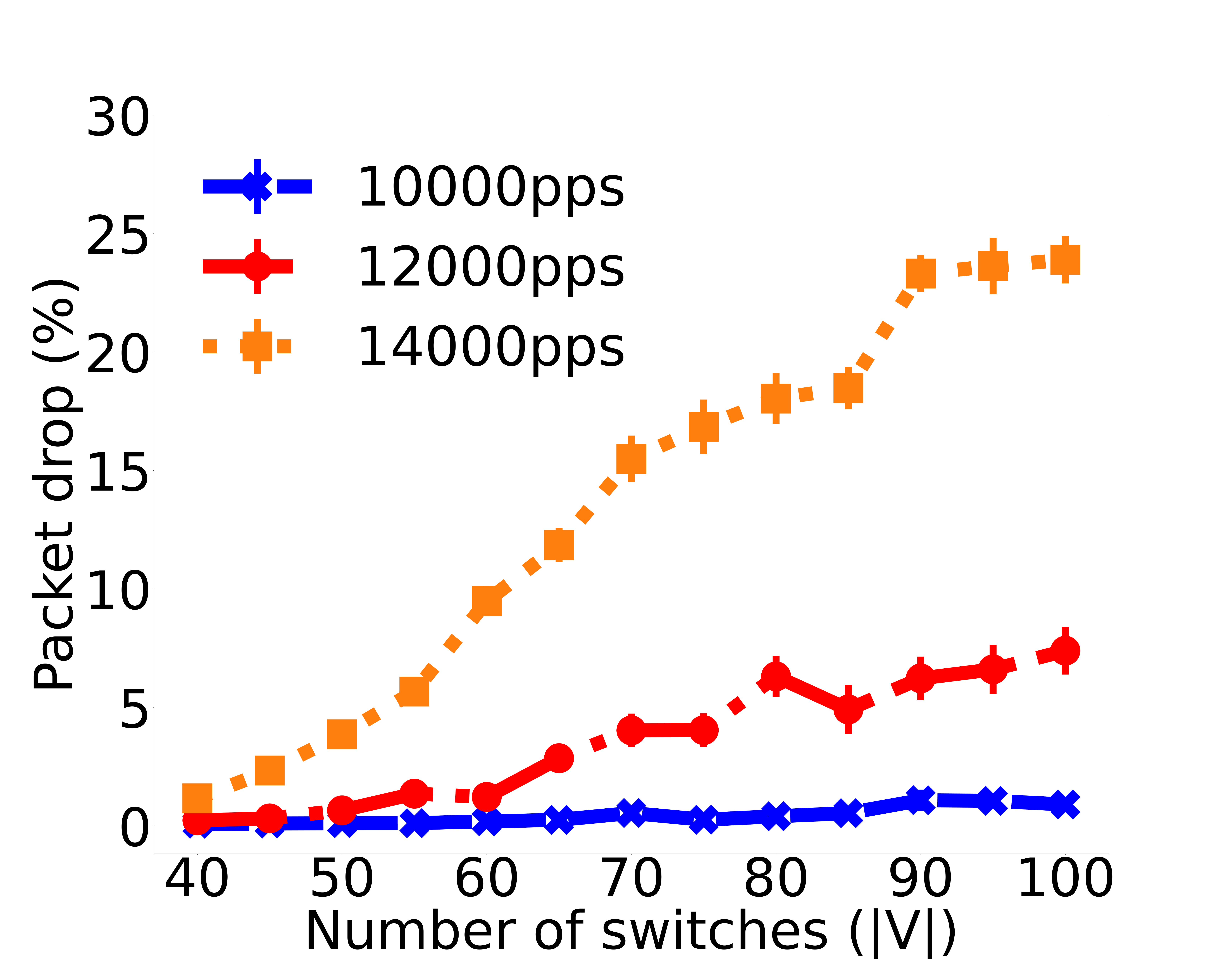}
    \caption{Data recovery disabled}
    \label{fig:sim-vertices-bt-packet-loss}
  \end{subfigure}
  \begin{subfigure}{0.49\columnwidth}
    \includegraphics[width=\textwidth]{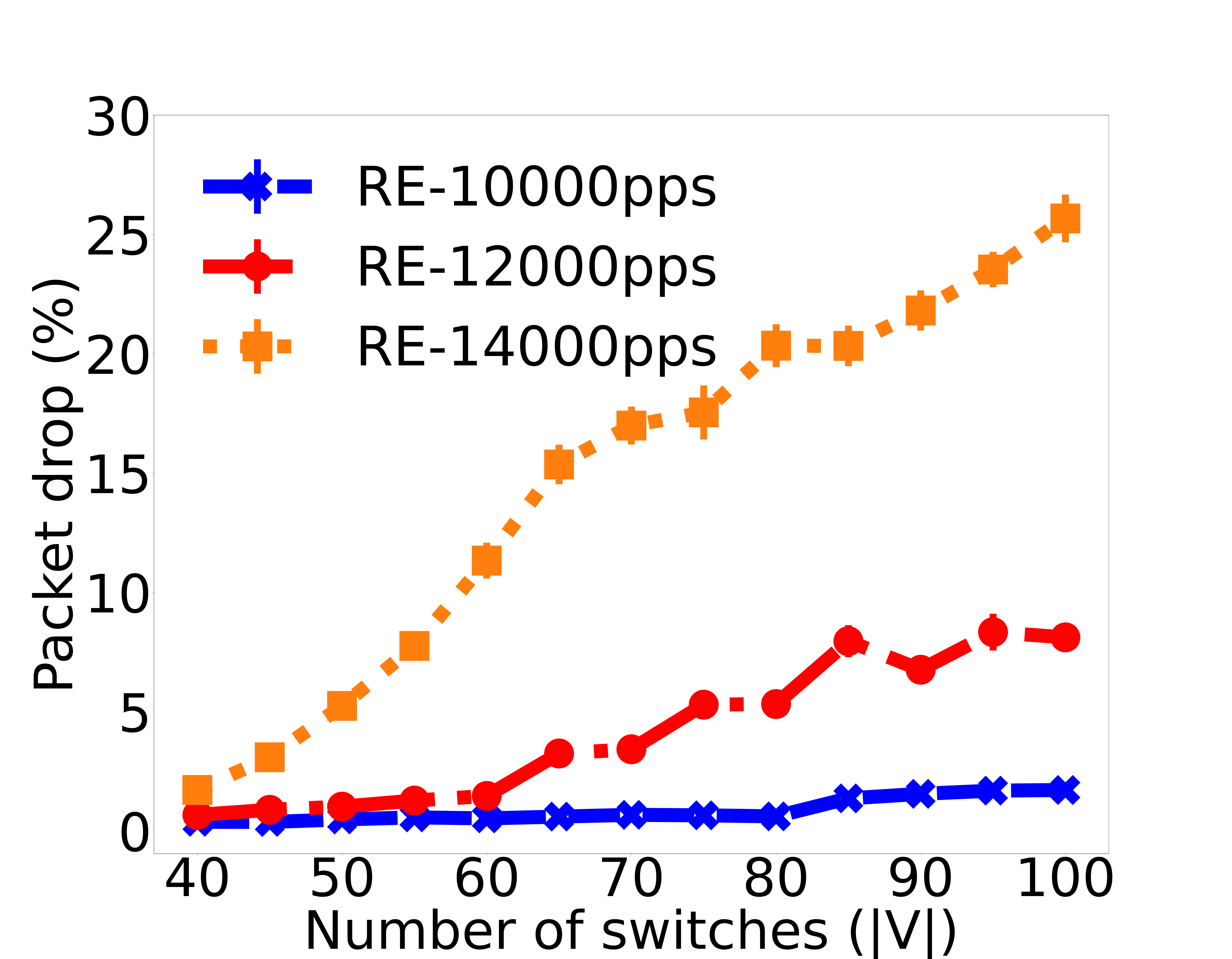}
    \caption{Data recovery enabled}
    \label{fig:sim-vertices-bt-robust-packet-loss}
  \end{subfigure}
  \caption{The ratio of packet loss for increasing network size}
  \label{fig:sim-vertices-bt-packet-losses}
\end{figure}

Before measuring the performance of our monitoring scheme, we first explain the impact of network congestion due to switch overutilization under different amounts of traffic loads s.t. 10000-14000 packets per second~(pps). Fig.~\ref{fig:sim-vertices-bt-packet-losses} shows the ratio of packet loss with and without data recovery. Both measurements are collected under active INT probing.

In both figures, the switches under 14000pps load reach 100\% packet processing utilization and then we observe packet drops. For 10000pps, the utilization stays between 85-92\% and thus it results in nearly 0\% drop. As it is shown in Fig.~\ref{fig:sim-vertices-bt-robust-packet-loss}, there is a 1-2\% increase in packet loss when the data recovery is enabled. The reason is, INT probes are stored on every switch with a limited queue size for $t_f$ seconds. Besides, probe feedback packets also circulate as additional control traffic. Both introduce a small control overhead and are also affected by the packet drops. Selecting a smaller $t_f$ can decrease the congestion in the packet queue but may trigger more recovery probes.

Note that those results are highly dependent on the processing power of the switches in the emulation environment, which also depends on the resources of the system hosting the emulation. Therefore, the results in Fig.~\ref{fig:sim-vertices-bt-packet-losses} can differ in a physical test bed. However, they indicate the baseline \emph{low} and \emph{high} traffic loads, i.e., 10000pps and 14000pps, triggering packet drops in our setup and thus consistently used for the comparison of different telemetry collection approaches.

\subsubsection{Disabled Data Recovery Module}

In this part, we compare our monitoring mechanism~(GPINT) with INT-Path~\cite{INTPath}~(Euler) and SNMP in terms of (i) the latency in telemetry collection and (ii) the failed telemetry collection sessions under low (10000pps) and high (14000pps) background traffic for increasing network size ($|V|$). The first evaluation metric indicates the required time conclude a measurement session, i.e., the elapsed time between sending the first probe and receiving the last. The latter metric indicates how often the monitoring module fails to collect \emph{all} probes, which can occur even at a loss of a single probe. Additionally, it also shows how often INT probes are lost during collection sessions. We do not enable the data recovery module for these experiments to measure the baseline performance of the respective approaches. Accordingly, we have not injected any link degradation yet and all packet losses stem from network congestion, i.e., drops in packet queues due to excessive load.

\begin{figure}[t]
  \begin{subfigure}{0.49\columnwidth}
    \includegraphics[width=\columnwidth]{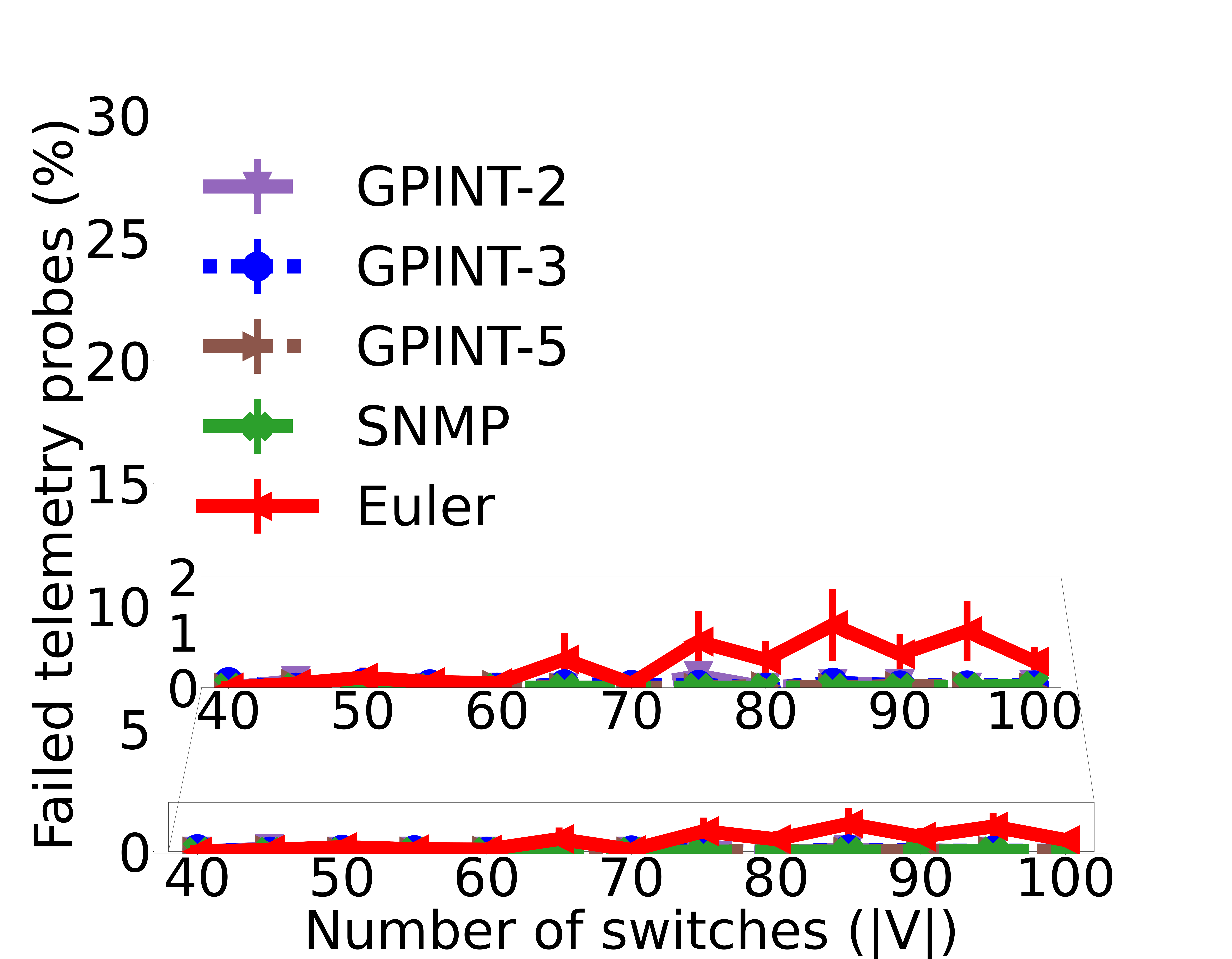}
    \caption{$10000$pps.}
    \label{fig:sim-vertices-bt-failed-paths-low}
  \end{subfigure}
  \begin{subfigure}{0.49\columnwidth}
    \includegraphics[width=\columnwidth]{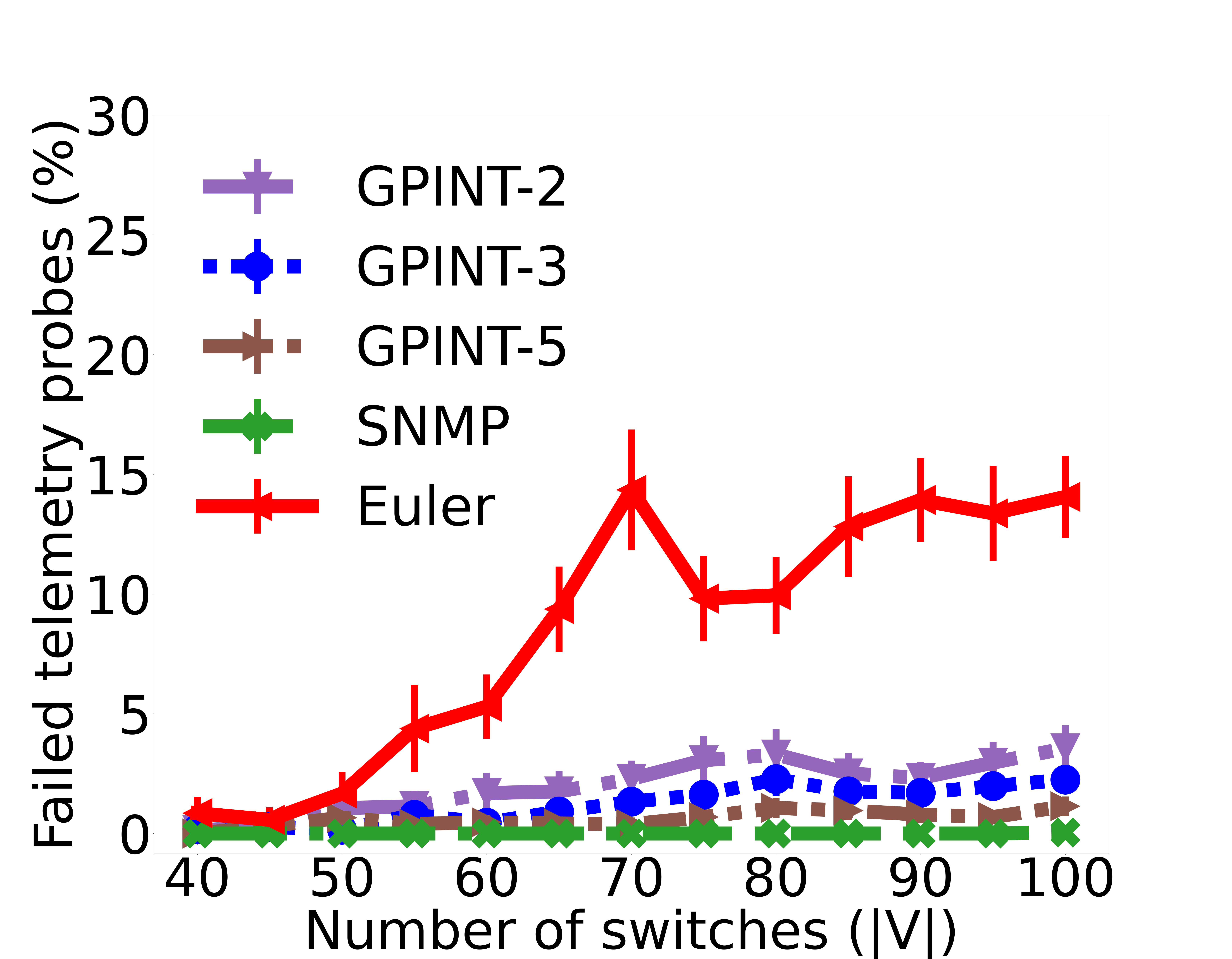}
    \caption{$14000$pps.}
    \label{fig:sim-vertices-bt-failed-paths-high}
  \end{subfigure}
  \caption{The ratio of failed telemetry probes to the number of generated probes.}
  \label{fig:sim-vertices-bt-failed-paths}
  \vspace{-4mm}
\end{figure}

A lost INT probe hinders to obtain full-network visibility and renders a telemetry collection session unsuccessful. However, it is still important to understand the exact ratio of packet drops, which can be higher than the ratio of failed sessions. Fig.~\ref{fig:sim-vertices-bt-failed-paths} shows the percentage of lost probes. In Fig.~\ref{fig:sim-vertices-bt-failed-paths-low}, the probe packet losses do not exceed 2\% under low load. Meanwhile, under high load, up to 15\% of probe packets deployed by Euler fail to arrive at the orchestrator as depicted in Fig.~\ref{fig:sim-vertices-bt-failed-paths-high}. On the other hand, the losses do not exceed 5\% for GPINT variations. The difference between GPINT and Euler shows the importance of balanced path generation outlined in our previous work~\cite{drcn_gpint}. We refer to this figure in the next sessions as well to discuss the performance of the recovery module comparatively.

\begin{figure}[h]
  \begin{subfigure}{0.49\columnwidth}
    \includegraphics[width=\columnwidth]{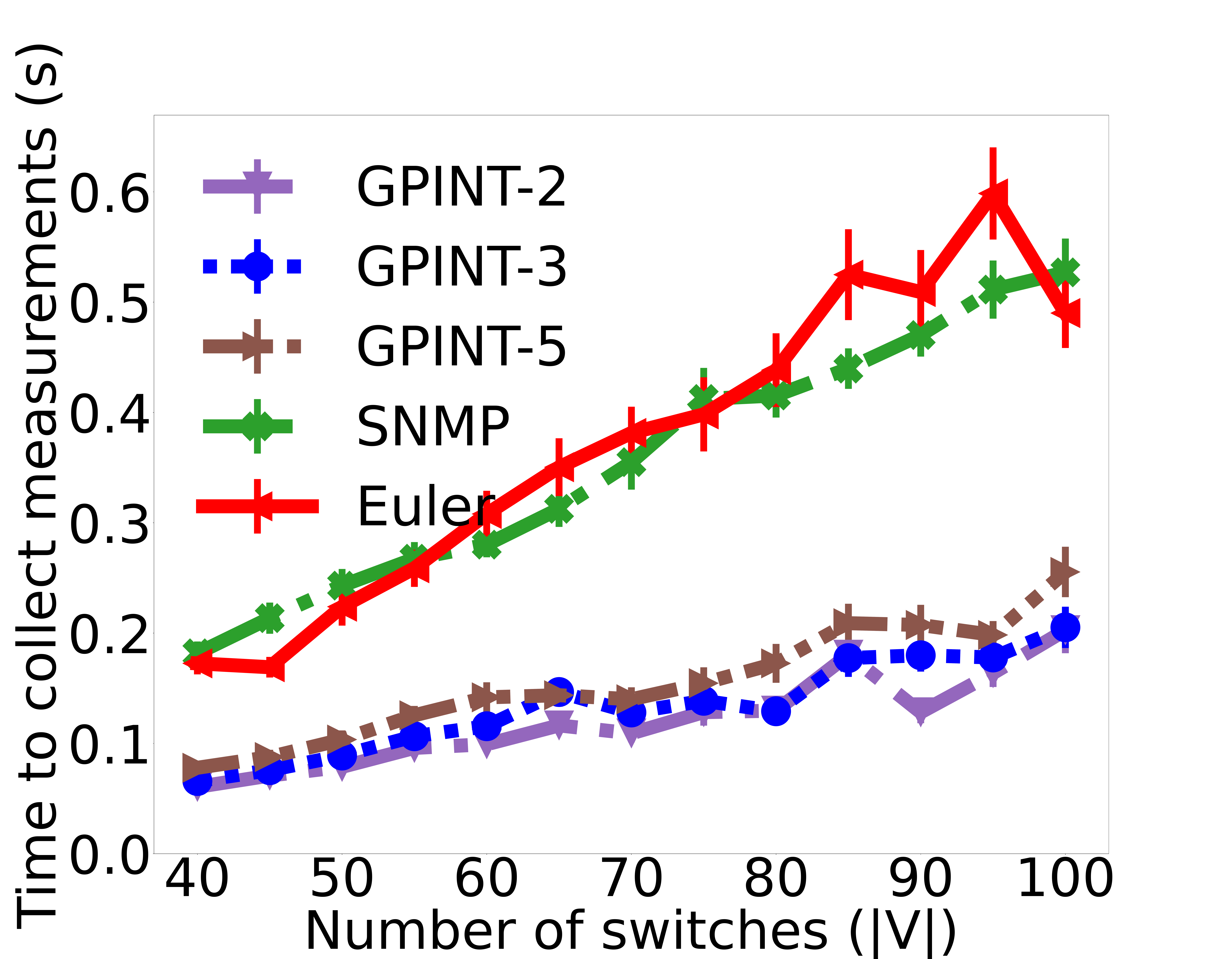}
    \caption{Elapsed time}
    \label{fig:sim-vertices-bt-elapsed-low}
  \end{subfigure}
  \begin{subfigure}{0.49\columnwidth}
    \includegraphics[width=\columnwidth]{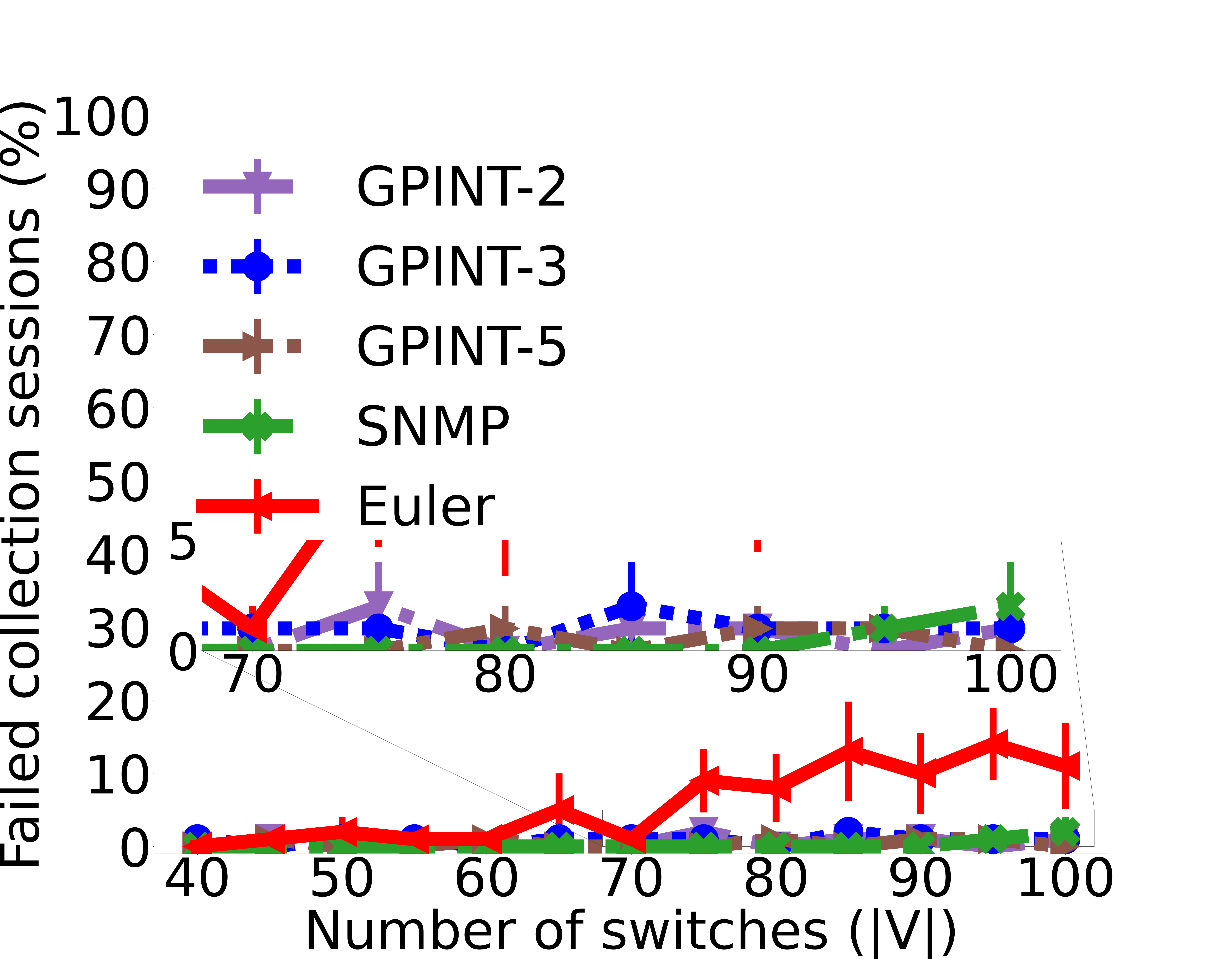}
    \caption{Failed collection sessions ($\%$)}
    \label{fig:sim-vertices-bt-failed-int-low}
  \end{subfigure}
  \caption{Data recovery disabled under low traffic load}
\end{figure}

Fig.~\ref{fig:sim-vertices-bt-elapsed-low} shows the latency in telemetry collection under low load. 
All GPINT approaches (with a different number of INT paths) results in the least latency as it deploys balanced disjoint paths with similar lengths. Euler, however, causes a substantial collection time due to considerably longer paths and accumulated queueing delays. Further discussion on such a delay accumulation can be found in our previous work~\cite{drcn_gpint}. Lastly, although SNMP is not affected by the probe queueing delay, it does not scale well given that the orchestrator in this experiment can only deploy 12 probes concurrently. Hence, SNMP spends the majority of time probing switches on larger scales.

Fig.~\ref{fig:sim-vertices-bt-failed-int-low} shows the percentage of failed telemetry collection sessions due to at least one lost probe packet.  Here, we do not introduce link degradation yet and these failures occur only due to congestion on switches. In the figure, Euler shows up to 10\% packet loss as it generates longer INT paths that traverse the network in longer time and consequently is more likely subject to packet losses. GPINT with balanced paths shows consistent results with Fig.~\ref{fig:sim-vertices-bt-packet-loss}, where the whole network is affected by the congestion. Additionally, there might be occasional SNMP packet losses due to congestions observed on switches.

\begin{figure}[t]
  \begin{subfigure}{0.49\columnwidth}
    \includegraphics[width=\columnwidth]{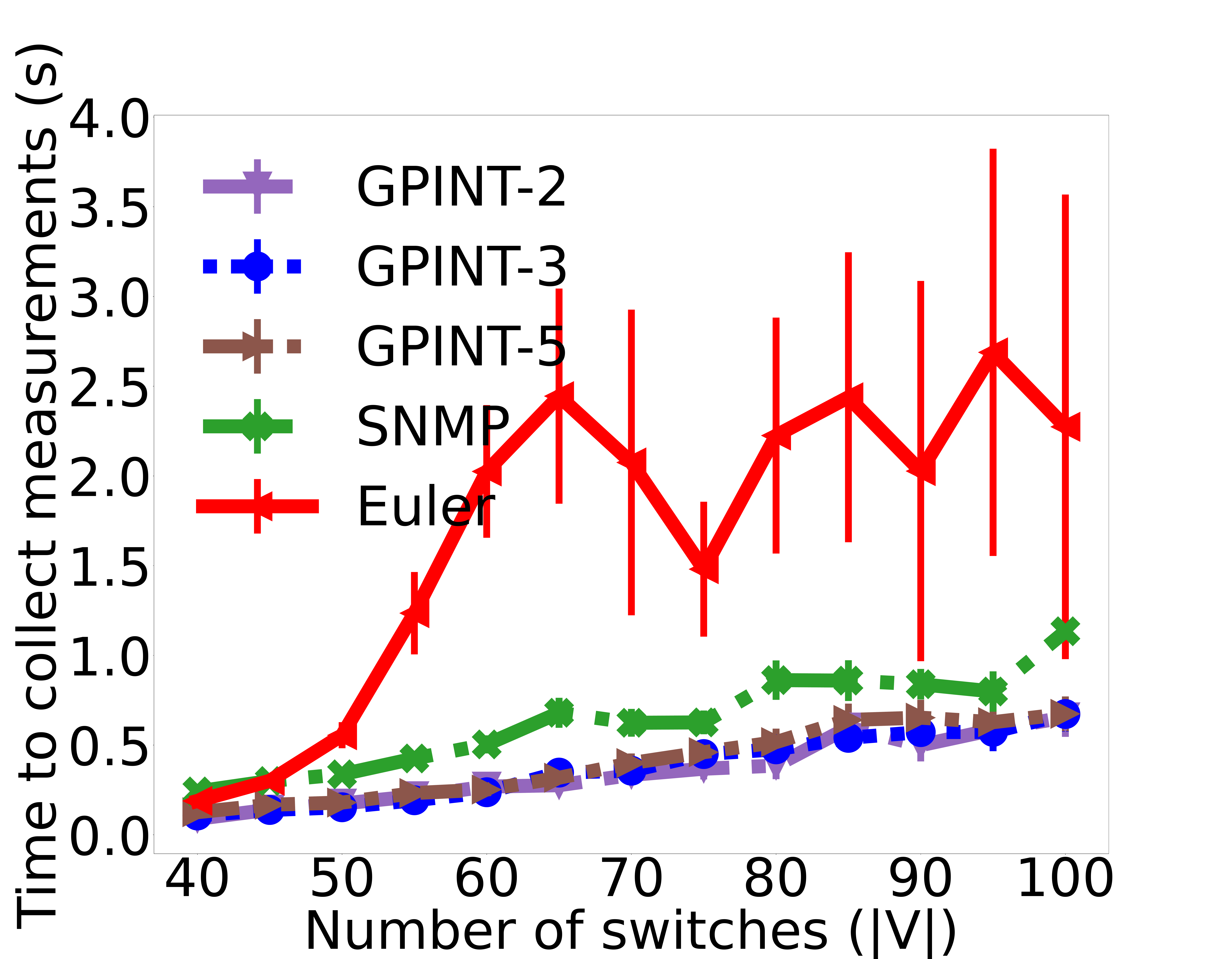}
    \caption{Elapsed time}
    \label{fig:sim-vertices-bt-elapsed-high}
  \end{subfigure}
  \begin{subfigure}{0.49\columnwidth}
    \includegraphics[width=\columnwidth]{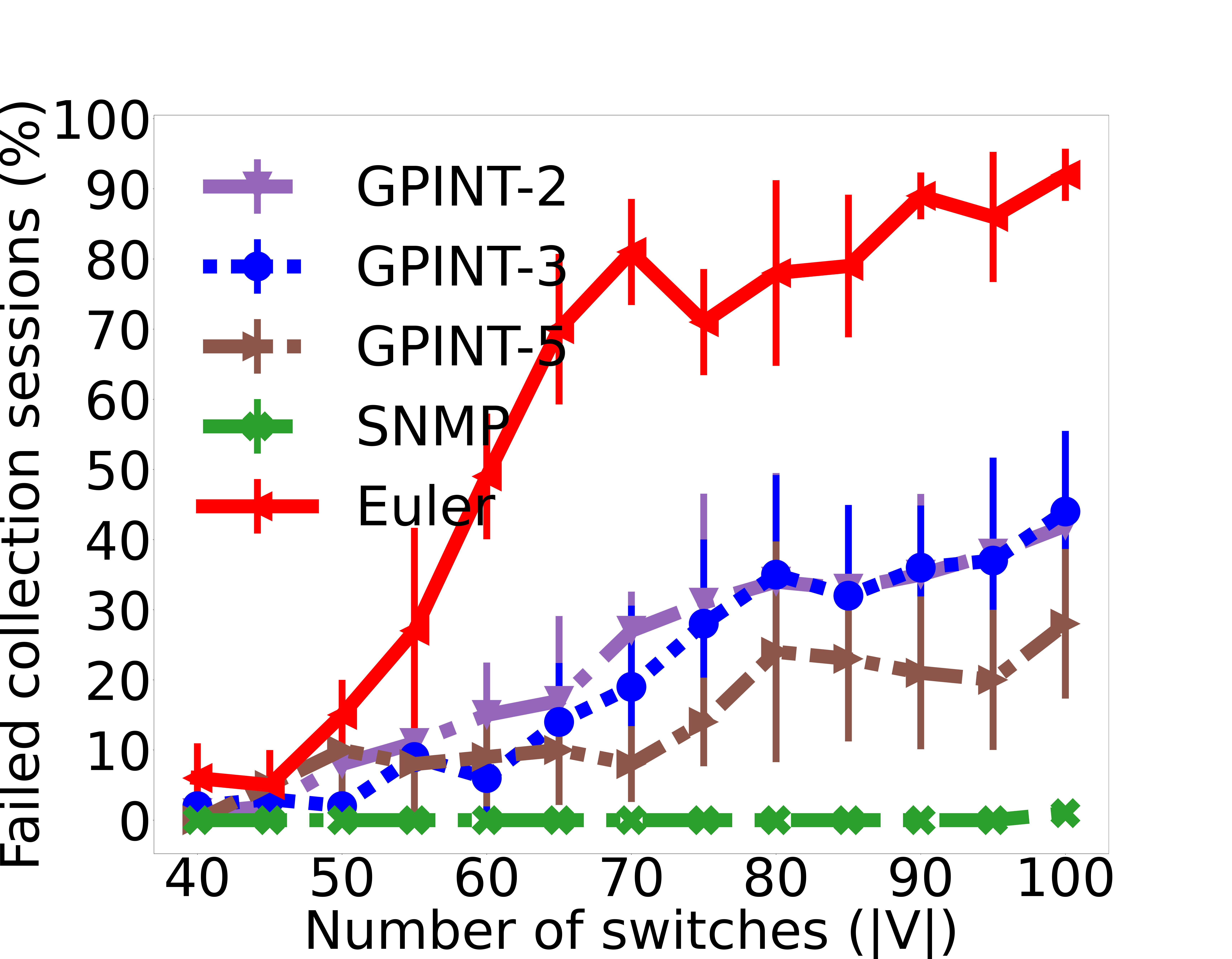}
    \caption{Failed collection sessions ($\%$)}
    \label{fig:sim-vertices-bt-failed-int-high}
  \end{subfigure}
  \caption{Data recovery disabled under high traffic load}
  \label{fig:sim-vertices-bt-high}
\end{figure}

Fig.~\ref{fig:sim-vertices-bt-high} shows the results of the same metrics under high traffic load. We observe an increase in telemetry collection latency for all approaches in Fig.~\ref{fig:sim-vertices-bt-elapsed-high}. As the load in the network increases, the gap between GPINT variations decreases due to more balanced paths and less overhead on switches, whose analysis is covered in our previous work~\cite{drcn_gpint}.
Whereas Euler shows exponential growth in delays up to 60 switches and then significant deviations in results due to very frequent packet losses and thus failed telemetry sessions as observed in Fig.~\ref{fig:sim-vertices-bt-failed-int-high}. GPINT, on the other hand, has much less loss that is under 40\% even for the largest network size. Moreover, as the $k$ increases, INT paths get shorter. As a result, each probe traverses the network in a shorter time and is less likely to experience a dropped probe. \\
\textbf{Impact of Link Degradation:} We also introduce degraded links, which are prone to probabilistic packet losses, and evaluate how INT monitoring performs without any data recovery mechanism. To emulate that, we configure the link properties of Mininet to set a 5-20\% packet-drop probability~(PDP) to each link between switches under high load.  Proportional to PDP, each packet has a certain probability of being lost during the propagation. We compare GPINT with SNMP for the evaluation.

\begin{figure}[t]
  \begin{subfigure}{0.49\columnwidth}
    \includegraphics[width=\columnwidth]{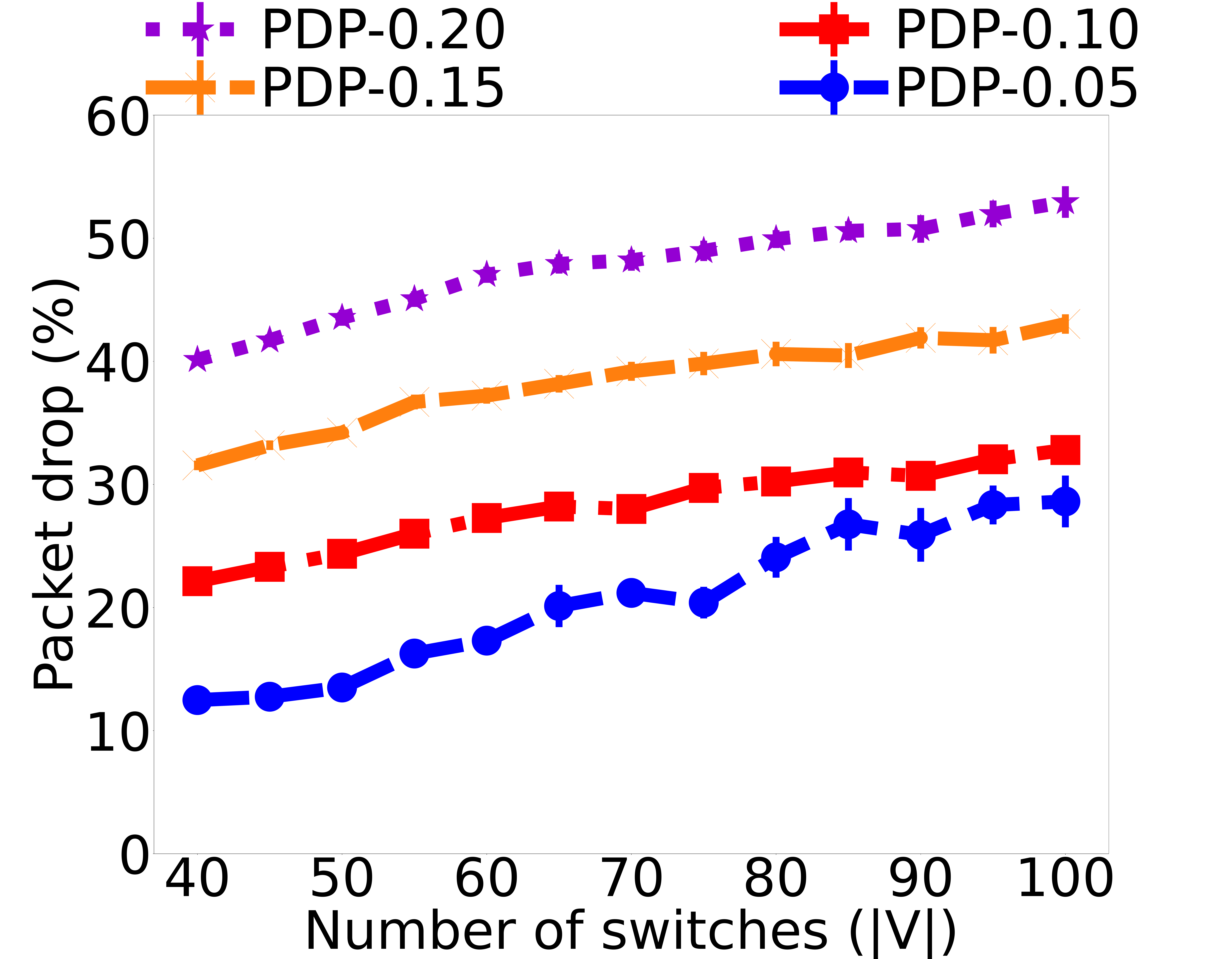}
    \caption{Background traffic loss}
    \label{fig:sim-vertices-bt-packet-loss-failure}
  \end{subfigure}
  \begin{subfigure}{0.49\columnwidth}
    \includegraphics[width=\columnwidth]{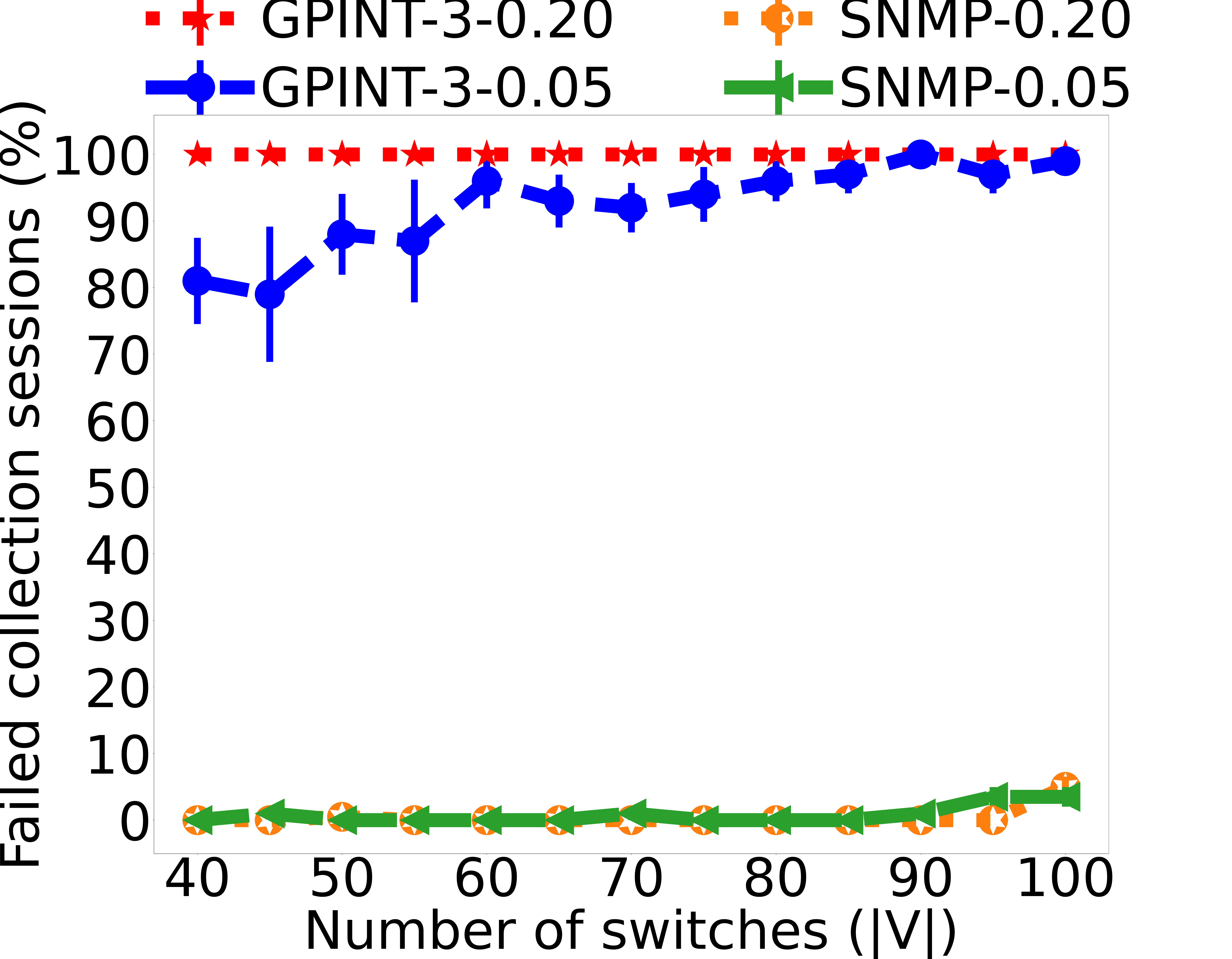}
    \caption{Failed collection sessions ($\%$)}
    \label{fig:sim-vertices-bt-failed-int-failure}
  \end{subfigure}
\center{
  \begin{subfigure}{0.49\columnwidth}
    \includegraphics[width=\columnwidth]{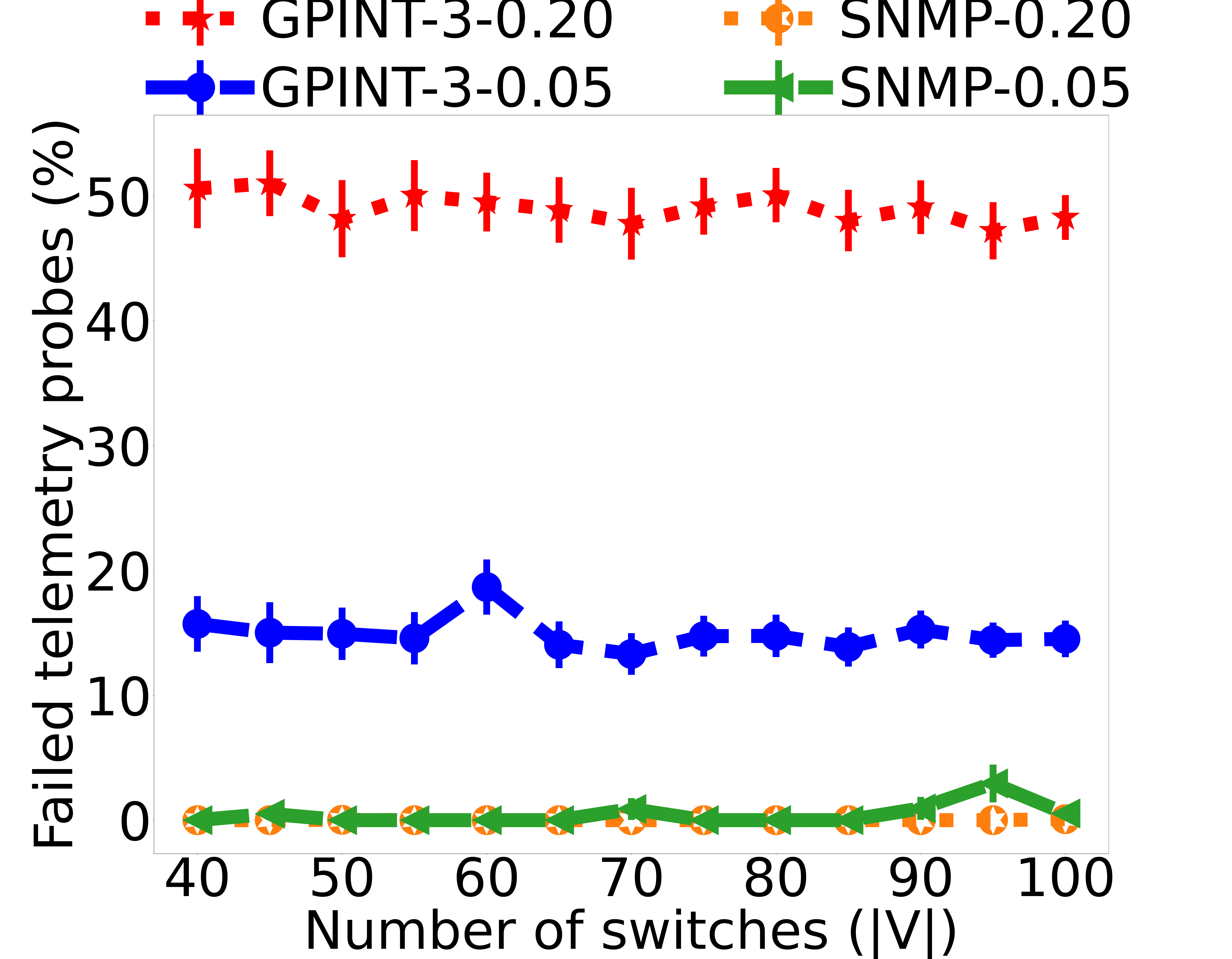}
    \caption{Failed probe packets ($\%$)}
    \label{fig:sim-vertices-bt-failed-paths-failure}
  \end{subfigure}
}
  \caption{Data recovery disabled under high load with link degradation}
  \label{fig:sim-vertices-bt-traffic-report-failure}
\end{figure}

Fig. \ref{fig:sim-vertices-bt-traffic-report-failure} shows how link degradation affects both traffic and INT probe losses.
We do not observe much of a traffic loss increase on 5-10\% PDP (shown as a ratio of 0.05 and 0.10) compared to Fig.~\ref{fig:sim-vertices-bt-packet-loss}.
However, after 10\%, we see that the background traffic losses can reach up to 40\% and 55\% for 15\% and 20\% PDP, respectively.
In Fig. \ref{fig:sim-vertices-bt-failed-int-failure}, it is seen that even 5\% PDP hinders consistent measurement deliveries as the failures in concluding a measurement session increase from 45\% to 80\% compared to Fig.~\ref{fig:sim-vertices-bt-failed-int-high}.
The probe packet losses take a great leap from 2.5\% (Fig.~\ref{fig:sim-vertices-bt-failed-paths-high}) to 20\% as depicted in Fig.~\ref{fig:sim-vertices-bt-failed-paths-failure}.
Hence, even subtle instabilities in the network can cause significant telemetry losses and degrade the ability of the monitoring module to reach accurate decisions using INT-based monitoring frameworks. SNMP also suffers subtle packet drops due to the congestion on the switches.

Eventually, these results show that a data recovery mechanism should be deployed for the probe-based INT, regardless of the probe generator. In the next section, we show the improvements that our data recovery mechanism introduces.

\subsubsection{Enabled Data Recovery Module} \label{sec:results-data-recovery}

In this section, we enable the data recovery module for all INT-Path approaches and measure the telemetry collection latency, the failed telemetry sessions, and the ratio of loss in INT probes, as well as the additionally generated recovery paths. Those recovery paths indicate how many probes the recovery module emits through the new INT paths to finalize a measurement session in case of a packet loss. It is calculated as $\frac{100\times |P_R|}{|P|}$, where $P_R$ is the set of generated recovery paths and $P$ is the set of initially generated paths by the \textit{probe generator}.
During the recovery, each switch needs to store a number of INT packets for $t_f$ seconds. The size of an INT probe is up to 2200~bytes according to the given packet layout.
Accordingly, $c$ concurrent telemetry collection sessions require $2200 \times c$ bytes for all INT probes to be stored on a switch. In this work, we use $c = 1$ for all experiments. However, when INT paths are not disjoint, a switch may store multiple probes at a time. In our previous work, we analyze intersection values of both Euler and GPINT and show that GPINT minimizes the number of intersections, hence the required space~\cite{drcn_gpint}.

\begin{figure}[h]
  \begin{subfigure}{0.49\columnwidth}
    \includegraphics[width=\columnwidth]{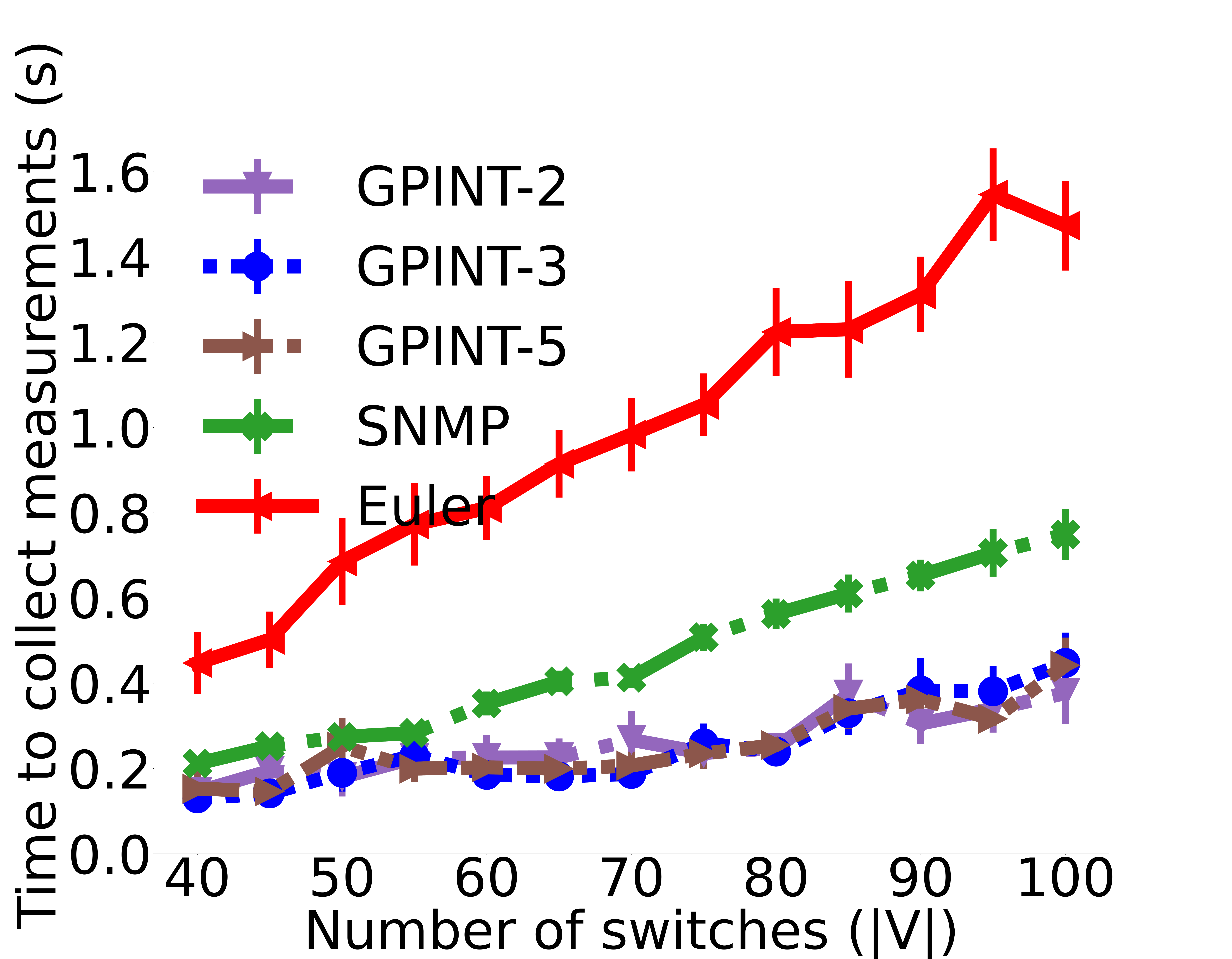}
    \caption{Elapsed time}
    \label{fig:sim-vertices-bt-robust-elapsed-low}
  \end{subfigure}
  \begin{subfigure}{0.49\columnwidth}
    \includegraphics[width=\columnwidth]{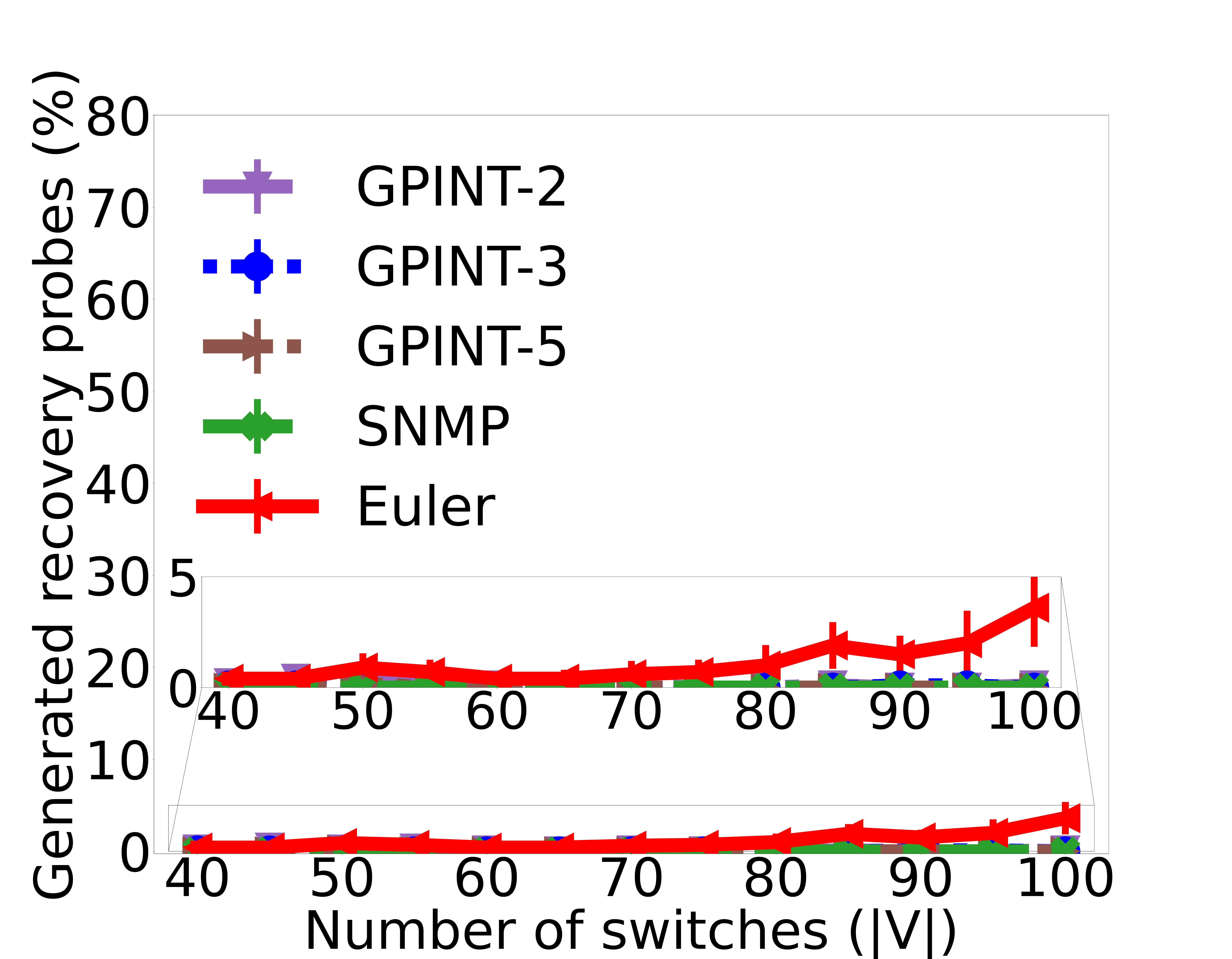}
    \caption{The ratio of recovery paths}
    \label{fig:sim-vertices-bt-robust-add-low}
  \end{subfigure}
  \caption{Data recovery enabled under low traffic load}
\end{figure}

In Fig.~\ref{fig:sim-vertices-bt-robust-elapsed-low}, we measure the telemetry collection latency and the additionally generated recovery paths under low load and with the enabled data recovery module.
We observe an extra delay of at most 0.15 seconds for GPINT variations, and SNMP compared to Fig.~\ref{fig:sim-vertices-bt-elapsed-low}.
For both, the ratio of generated recovery paths shown in Fig.~\ref{fig:sim-vertices-bt-robust-add-low} also matches with the probe packet losses depicted in Fig.~\ref{fig:sim-vertices-bt-failed-paths-low}. It indicates that the overhead for data recovery in terms of extra deployed paths is limited. On the other hand, Euler has up to 1.6 seconds delay nearly two times higher than the results in Fig.~\ref{fig:sim-vertices-bt-elapsed-low} without data recovery. Moreover, the ratio of additionally generated recovery paths is about 5-6\%, which is greater than the observed probe packet losses depicted in Fig.~\ref{fig:sim-vertices-bt-failed-paths-low}.
The reason is, the induced overhead by the recovery module causes additional packet losses, especially in the case of long INT paths of Euler, and thus degrades its performance as also discussed in the previous section.

\begin{figure}[t]
  \begin{subfigure}{0.49\columnwidth}
    \includegraphics[width=\columnwidth]{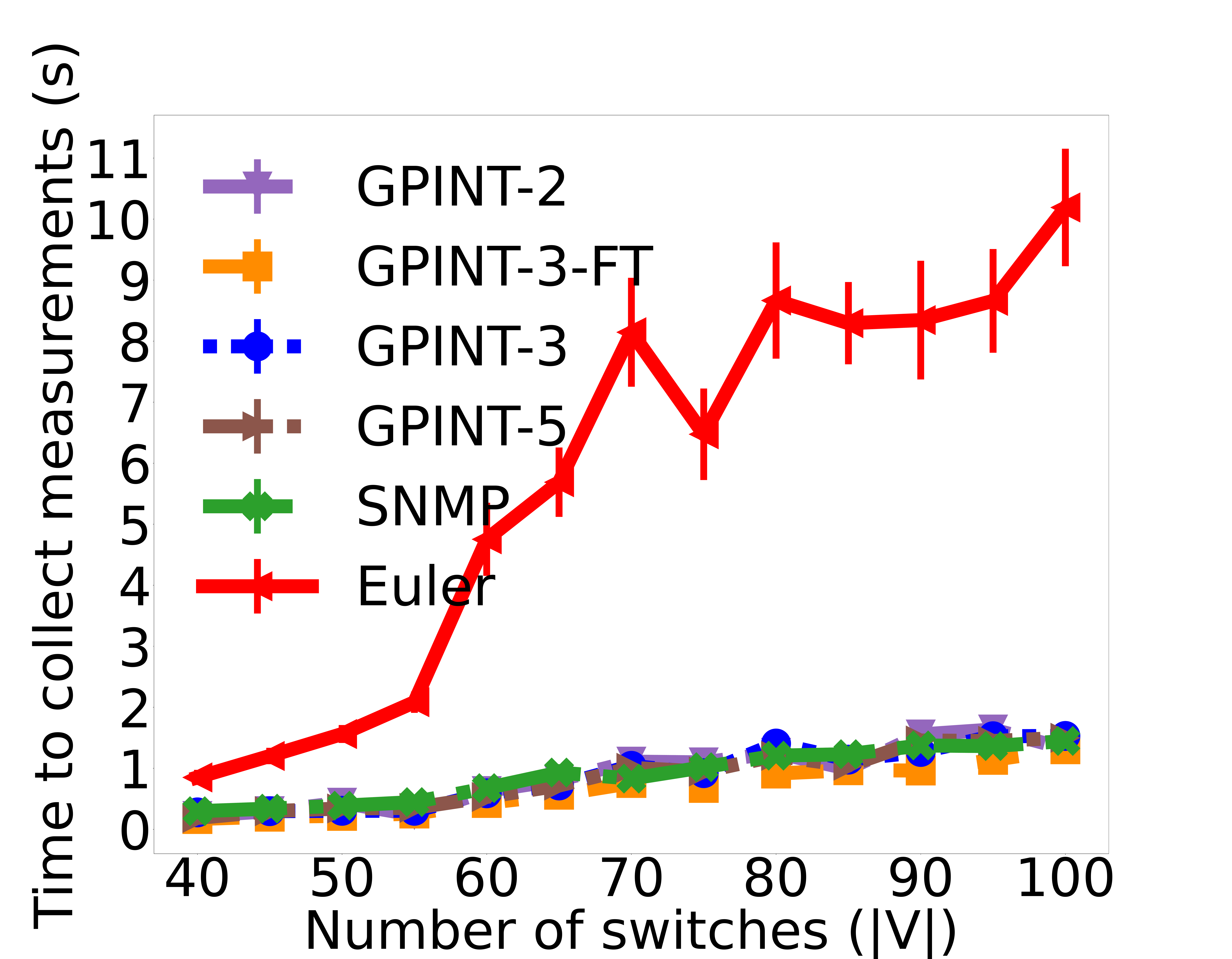}
    \caption{Elapsed time}
    \label{fig:sim-vertices-bt-robust-elapsed-high}
  \end{subfigure}
  \begin{subfigure}{0.49\columnwidth}
    \includegraphics[width=\columnwidth]{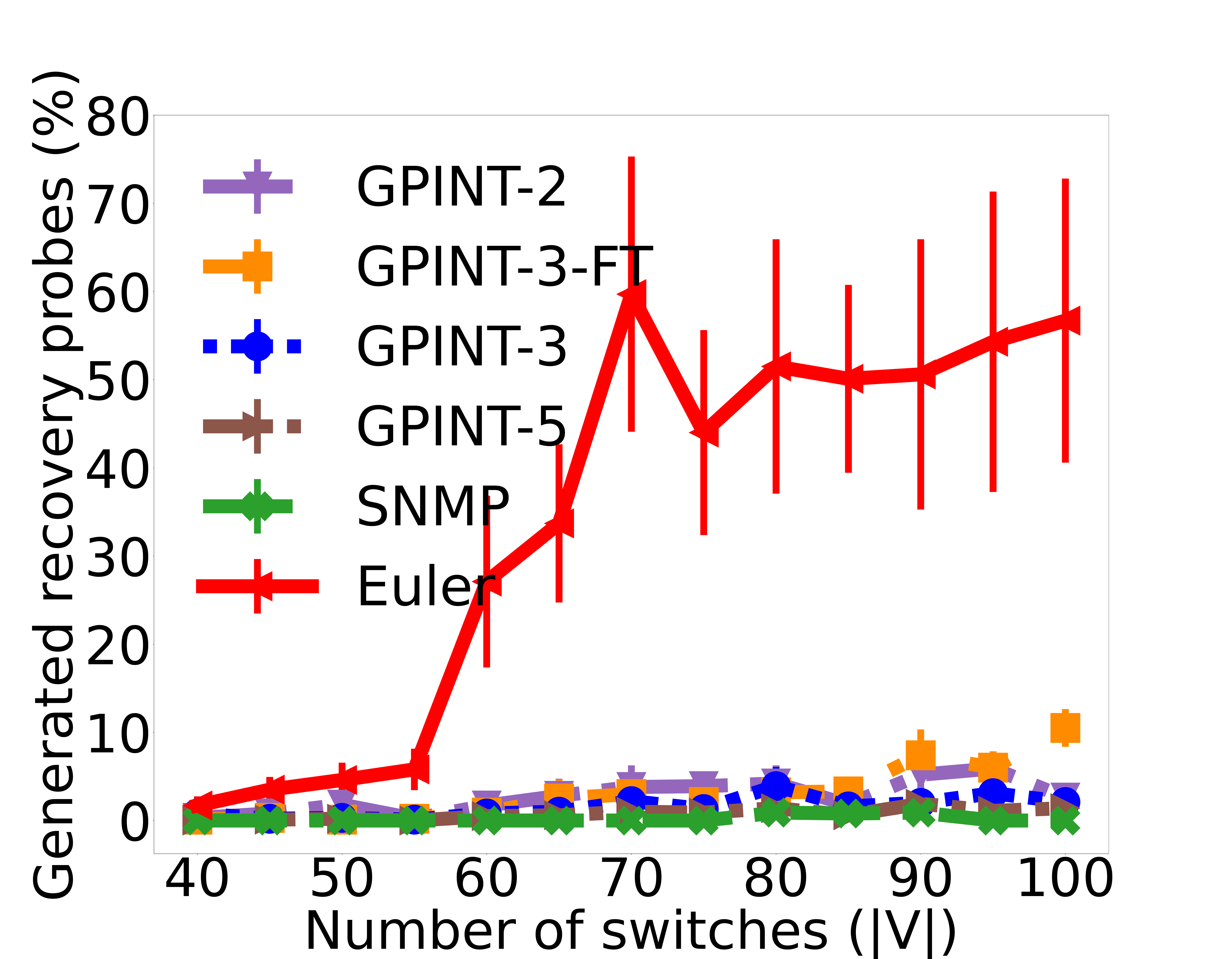}
    \caption{The ratio of recovery paths}
    \label{fig:sim-vertices-bt-robust-add-high}
  \end{subfigure}
  \caption{Data recovery enabled under high traffic load}
  \label{fig:sim-vertices-bt-robust-high}
\end{figure}

Lastly, Fig.~\ref{fig:sim-vertices-bt-robust-high} presents the results for latency and additionally generated paths under high load. Here, we employ another variation of GPINT, a fine-tuned GPINT-FT, to trigger recovery process much faster selecting $t_f=0.4$~seconds $r_a=2$ with the cost of generating more probe packets. In the figure, all GPINT variations perform worse by 0.1 seconds than SNMP in Fig.~\ref{fig:sim-vertices-bt-robust-elapsed-high}, except GPINT-3-FT, which is configured with more aggressive design parameters.
Furthermore, they require slightly more time than their performance under low load to recover from packet losses occurring more often. Comparing Fig.~\ref{fig:sim-vertices-bt-failed-paths-high} and Fig.~\ref{fig:sim-vertices-bt-robust-add-high}
reveals that the recovery module generates almost the same number of paths as failed probe packets, indicating that a recovery packet does not or rarely fail for GPINT.
Additionally, GPINT-3-FT generates slightly more probes due to its aggressive design parameters.
On the other hand, there is a considerable difference for Euler between failed and deployed recovery probes. This is a clear indication that the recovery probes can also fail in the network. The main reason behind this behavior is the Euler generating long and unbalanced paths, which spend more time in the network and are subject to higher failure chances. Accordingly, the elapsed time to collect telemetry for Euler can take up to 10 seconds.

\begin{figure}[t!]
  \begin{subfigure}{0.49\columnwidth}
    \includegraphics[width=\columnwidth]{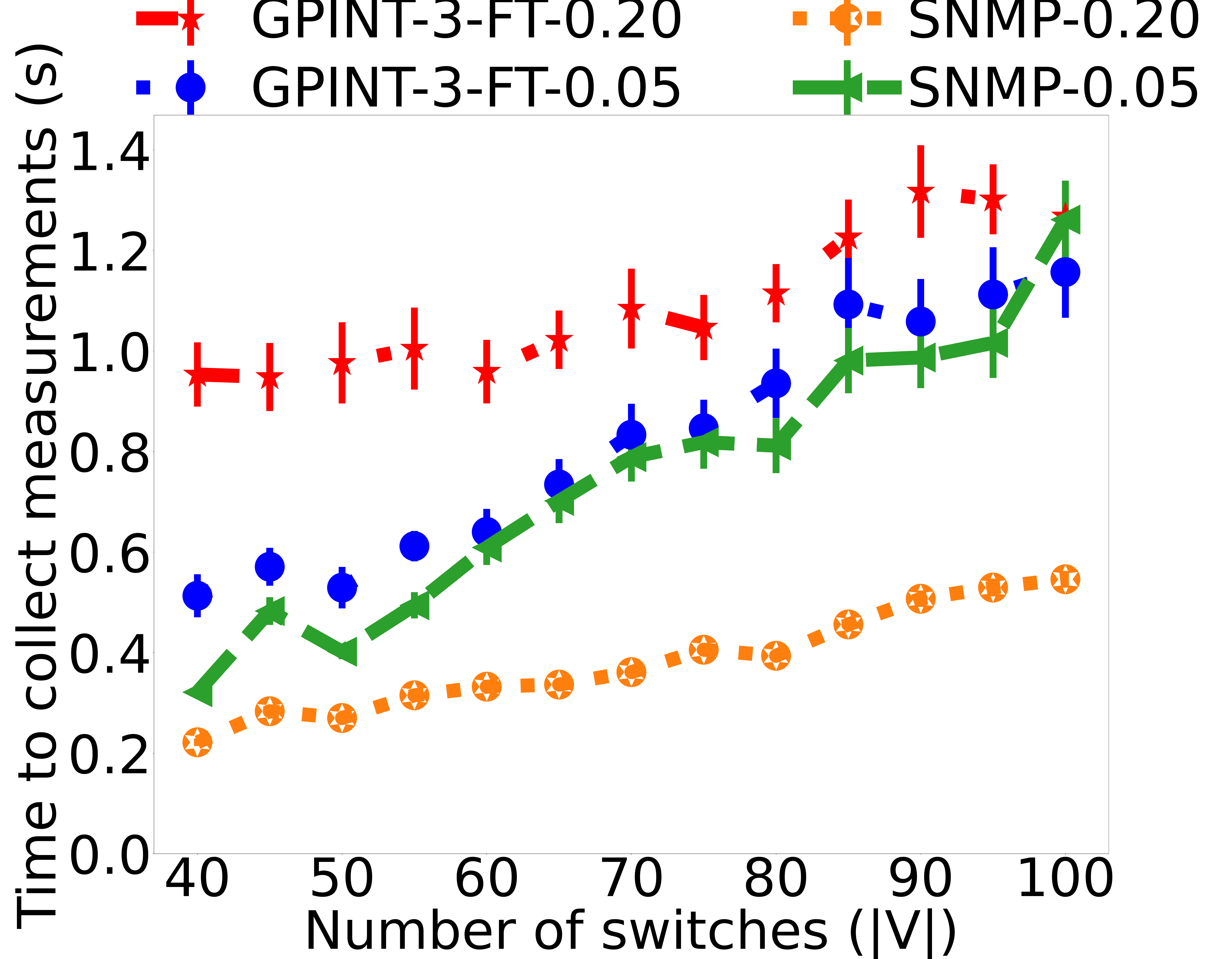}
    \caption{Elapsed time}
    \label{fig:sim-vertices-bt-robust-elapsed-failure}
  \end{subfigure}
  \begin{subfigure}{0.49\columnwidth}
    \includegraphics[width=\columnwidth]{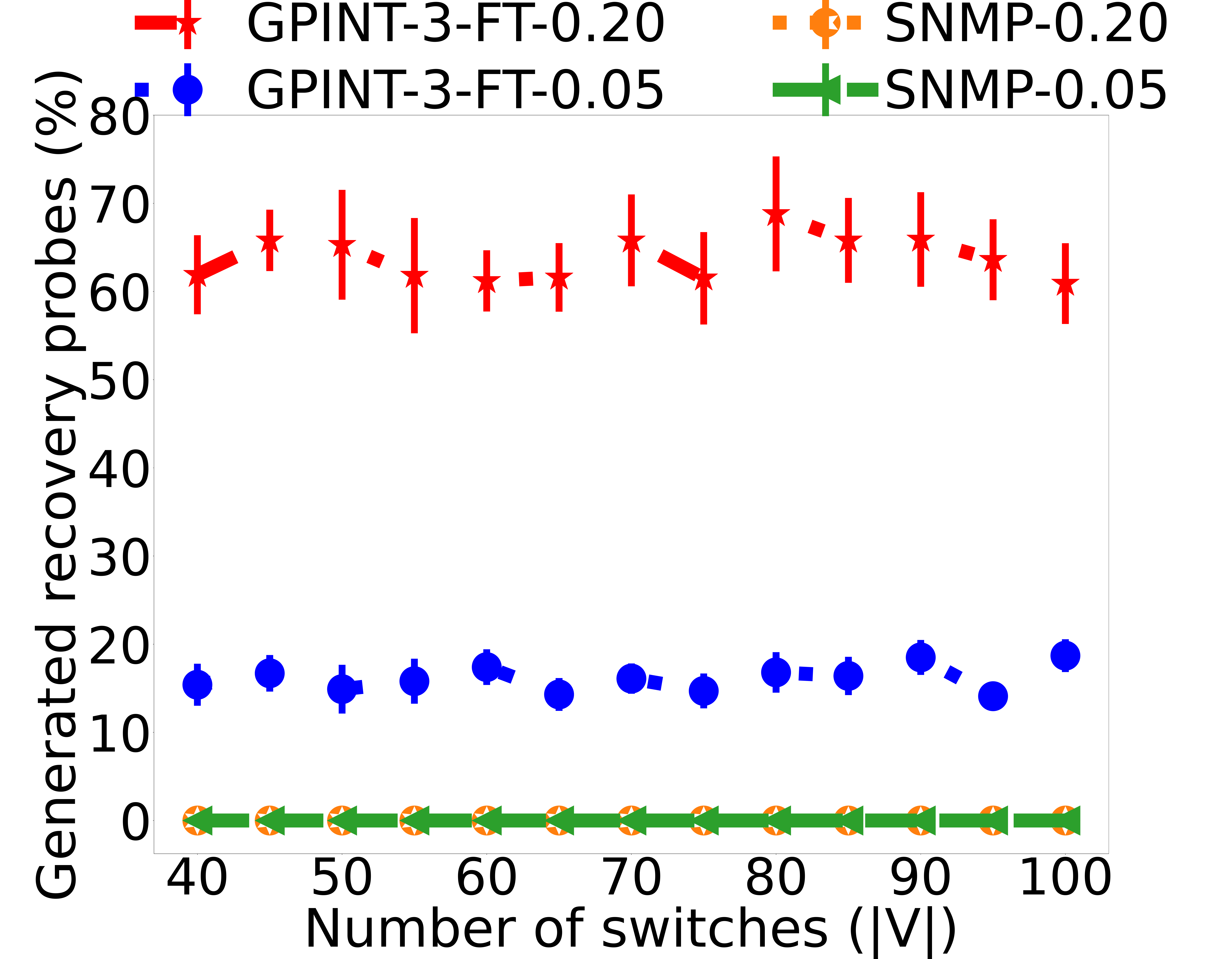}
    \caption{The ratio of recovery probes}
    \label{fig:sim-vertices-bt-robust-add-failure}
  \end{subfigure}
  \caption{Data recovery enabled under high traffic load with link degradation}
  \label{fig:sim-vertices-bt-robust-failure}
\end{figure}

These results show that the data recovery module has a negligible low overhead.
They also reveal the importance of the balanced INT paths as they can almost seamlessly be recovered.
Additionally, a balanced path generation with shorter paths provides the best reliability as we observe the least packet failures for GPINT-5 in Fig.~\ref{fig:sim-vertices-bt-failed-paths-high}. However, it might be better to deploy a different data recovery mechanism for Euler considering long and unbalanced paths. \\
\textbf{Impact of Link Degradation:} As in the previous section, we evaluate the impact of link degradation in presence of the data recovery module. Fig.~\ref{fig:sim-vertices-bt-robust-failure} shows the evaluation of fine-tuned for GPINT-3 and SNMP under link degradations.
For GPINT, we observe a significant difference in telemetry collection latency between different link degradation probabilities, i.e., 0.05-0.20 representing 5-20\%.
It is also shown that the performance of GPINT-3-FT is similar for 5\% and 20\% PDP. This behavior can be explained as follows.
As PDP increases, the switches process fewer packets. The data load on the switches and packet loss due to the congestion decrease proportionally, and consequently INT probes spend less time on the packet queue and are being forwarded faster.
When a packet loss is detected under these circumstances, it is more likely due to link degradation.
Hence, switches update the controller more frequently on the whereabouts of INT probes so that the data recovery module can release additional probes via recovery paths close to endpoints rather than starting from the beginning of an INT path.

On the other hand, the performance of SNMP increases with an increasing PDP as switches become less congested. On 5\% PDP (0.05), the gap between GPINT and SNMP remains rather close. This indicates that under subtle packet losses, the benefits of INT can still be leveraged with the help of the data recovery module. However, as PDP increases, the gap between SNMP and GPINT widens. Accordingly, employing INT in challenging/harsh environments requires (i) careful path planning and (ii) proactive configuration of the data recovery module.

In Fig.~\ref{fig:sim-vertices-bt-robust-add-failure}, we depict the ratio of additionally generated recovery probes to initially sent INT probes.
In comparison to the failed telemetry probes depicted in Fig.~\ref{fig:sim-vertices-bt-failed-paths-failure}, the recovery module generates almost the same amount of recovery probes for 5\% PDP. For 20\% PDP, it generates 10-20\% more paths than the lost probe packets, which means that the deployed recovery packets also suffer from losses, and the module generates more probes to complete the telemetry collection sessions.
Regardless of these extra recovery packets, the total number of generated paths to conclude a measurement session is considerably smaller than probing every switch individually.
Given that GPINT-3 generates $\sim$25 INT probes to cover the whole network of 100 switches~\cite{drcn_gpint}, the total number of probes generated to obtain reliable results is $\sim$30 and $\sim$42 under 5\% and 20\% PDP, respectively. Accordingly, INT with a data recovery module offers better scalability compared to probing individual switches like SNMP, even under link degradations.

This experiment shows that the data recovery module can perform considerably well on networks that may experience unstable links, such as wireless networks.
By fine-tuning the configurable parameters that the module offers, one can adapt it to any condition. However, it still has its limitations that we discussed in previous analyses and works best with path generators that can generate balanced paths.

\section{Conclusion and Future Work} \label{sec:conclusion}

Network monitoring solutions should adapt to the increasing complexity of the modern networked systems. Especially from a scalability and flexibility perspective, the traditional solutions like SNMP and even recent technologies like SDN have certain drawbacks. In our previous work~\cite{drcn_gpint}, we leveraged network programmability with P4 to design a probe-based, distributed, and scalable monitoring scheme with an in-band network telemetry~(INT) framework. Here, we extend our work with a data recovery mechanism to guarantee a timely and reliable network telemetry collection routine. With this extension, we satisfy three main requirements for the design of a monitoring system namely, (i) full-network coverage, (ii) timeliness, and (iii) reliability, which are more extensively discussed in Section~\ref{sec:probdef}. For the evaluation, we implemented our proposed mechanism including all design modules, a telemetry collection protocol definition, and required packet structures in P4. We then compared it with SNMP and another state-of-the-art probe-based INT mechanism under different traffic loads and packet loss ratios by modeling link reliability in the emulation environment. First, our results reveal that INT is affected by any packet loss carrying telemetry information significantly as a single INT probe carries accumulated information for a partial network. Thus, we discuss the importance of the third requirement, reliability. We also show that when our data recovery mechanism is deployed, it is possible to overcome all of the probe losses even under high data traffic congestion and link degradation. Our proposal performs better both in terms of latency and control overhead compared to its opponents with increasing network size.

Lastly, we discussed the potential limitations of the data recovery module in the presence of degraded links, which are prone to packet loss. For instance, the reaction of the data recovery module is increasing under highly-degraded links with a high probability of packet loss. A better reaction time requires further analysis of our design parameters to find the optimal values for given network conditions. Another limitation is, that the data recovery module follows a single recovery strategy and does not adapt to the given conditions of the system. For instance, although a potentially congested switch can be easily identified, utilizing that information to find more suitable INT paths and embrace a more load-balanced telemetry scheme is not considered in the scope of this study. We focus on those limitations in our future work.


\ifCLASSOPTIONcaptionsoff
  \newpage
\fi



%
\balance
\bibliographystyle{ieeetr} 
\bibliography{references} 

\begin{thebibliography}{10}

\bibitem{iot_survey}
{Ola Salman and Imad Elhajj and Ali Chehab and Ayman Kayssi}, ``{IoT survey: An
  SDN and fog computing perspective},'' {\em Computer Networks}, vol.~143,
  pp.~221--246, 2018.

\bibitem{snmp}
J.~Case, M.~Fedor, M.~Schoffstall, and J.~Davin, ``{Simple network management
  protocol},'' tech. rep., STD 15, RFC 1157, SNMP Research, Performance Systems
  International, MIT, 1990.

\bibitem{netFlow}
B.~Claise, ``{Cisco Systems NetFlow Services Export Version 9},'' RFC 3954,
  2004.

\bibitem{sflow_techreport}
{P. Phaal, S. Panchen, N. McKee}, ``{InMon Corporation's sFlow: A Method for
  Monitoring Traffic in Switched and Routed Networks},'' RFC 3176, 2001.

\bibitem{traditional_monitoring_yang}
M.~Yang, Y.~Li, D.~Jin, L.~Zeng, X.~Wu, and A.~V. Vasilakos,
  ``{Software-Defined and Virtualized Future Mobile and Wireless Networks: A
  Survey},'' {\em Mobile Networks and Applications}, vol.~20, no.~1, pp.~4--18,
  2014.

\bibitem{traditional_monitoring_jammal}
M.~Jammal, T.~Singh, A.~Shami, R.~Asal, and Y.~Li, ``{Software defined
  networking: State of the art and research challenges},'' {\em Computer
  Networks}, vol.~72, pp.~74--98, 2014.

\bibitem{network_monitoring_sdn_review}
P.~{Tsai}, C.~{Tsai}, C.~{Hsu}, and C.~{Yang}, ``{Network Monitoring in
  Software-Defined Networking: A Review},'' {\em IEEE Systems Journal},
  vol.~12, no.~4, pp.~3958--3969, 2018.

\bibitem{traditional_scalability_pras}
A.~{Pras}, J.~{Schonwalder}, M.~{Burgess}, O.~{Festor}, G.~M. {Perez},
  R.~{Stadler}, and B.~{Stiller}, ``{Key research challenges in network
  management},'' {\em IEEE Communications Magazine}, vol.~45, no.~10,
  pp.~104--110, 2007.

\bibitem{sdn_distributed}
F.~{Bannour}, S.~{Souihi}, and A.~{Mellouk}, ``{Distributed SDN Control:
  Survey, Taxonomy, and Challenges},'' {\em IEEE Communications Surveys
  Tutorials}, vol.~20, no.~1, pp.~333--354, 2018.

\bibitem{sdn_wireless_sensor}
H.~I. {Kobo}, A.~M. {Abu-Mahfouz}, and G.~P. {Hancke}, ``{A Survey on
  Software-Defined Wireless Sensor Networks: Challenges and Design
  Requirements},'' {\em IEEE Access}, vol.~5, pp.~1872--1899, 2017.

\bibitem{sdn_monitoring_challenges}
A.~{Yassine}, H.~{Rahimi}, and S.~{Shirmohammadi}, ``{Software defined network
  traffic measurement: Current trends and challenges},'' {\em {IEEE
  Instrumentation \& Measurement Magazine}}, vol.~18, no.~2, pp.~42--50, 2015.

\bibitem{sdn_monitoring_sezer}
S.~{Sezer}, S.~{Scott-Hayward}, P.~K. {Chouhan}, B.~{Fraser}, D.~{Lake},
  J.~{Finnegan}, N.~{Viljoen}, M.~{Miller}, and N.~{Rao}, ``{Are we ready for
  SDN? Implementation challenges for software-defined networks},'' {\em IEEE
  Communications Magazine}, vol.~51, no.~7, pp.~36--43, 2013.

\bibitem{openflow}
N.~McKeown, T.~Anderson, H.~Balakrishnan, G.~Parulkar, L.~Peterson, J.~Rexford,
  S.~Shenker, and J.~Turner, ``{OpenFlow: Enabling Innovation in Campus
  Networks},'' {\em {SIGCOMM Comput. Commun. Rev.}}, vol.~38, pp.~69--74, 2008.

\bibitem{p4_review_kaur}
S.~Kaur, K.~Kumar, and N.~Aggarwal, ``{A review on P4-Programmable data planes:
  Architecture, research efforts, and future directions},'' {\em Comput.
  Commun.}, vol.~170, pp.~109--129, 2021.

\bibitem{programmable_review_bifulco}
R.~{Bifulco} and G.~{Rétvári}, ``A survey on the programmable data plane:
  Abstractions, architectures, and open problems,'' in {\em IEEE 19th
  International Conference on High Performance Switching and Routing (HPSR)},
  pp.~1--7, 2018.

\bibitem{p4_review_exhaustive}
E.~F. Kfoury, J.~Crichigno, and E.~Bou-Harb, ``{An Exhaustive Survey on P4
  Programmable Data Plane Switches: Taxonomy, Applications, Challenges, and
  Future Trends},'' 2021.

\bibitem{p4_paper}
P.~Bosshart, D.~Daly, G.~Gibb, M.~Izzard, N.~McKeown, J.~Rexford,
  C.~Schlesinger, D.~Talayco, A.~Vahdat, G.~Varghese, and D.~Walker, ``{P4:
  Programming Protocol-independent Packet Processors},'' {\em SIGCOMM Comput.
  Commun. Rev.}, vol.~44, no.~3, pp.~87--95, 2014.

\bibitem{programmable_review_kaljic}
{E. {Kaljic} and A. {Maric} and P. {Njemcevic} and M. {Hadzialic}}, ``{A Survey
  on Data Plane Flexibility and Programmability in Software-Defined
  Networking},'' {\em IEEE Access}, vol.~7, pp.~47804--47840, 2019.

\bibitem{int_paper}
C.~Kim, A.~Sivaraman, N.~Katta, A.~Bas, A.~Dixit, and L.~J. Wobker, ``{In-band
  network telemetry via programmable dataplanes},'' in {\em ACM SIGCOMM},
  vol.~15, 2015.

\bibitem{INTPath}
T.~{Pan}, E.~{Song}, Z.~{Bian}, X.~{Lin}, X.~{Peng}, J.~{Zhang}, T.~{Huang},
  B.~{Liu}, and Y.~{Liu}, ``{INT-Path}: {Towards Optimal Path Planning for
  In-band Network-Wide Telemetry},'' in {\em {Proc. of the IEEE INFOCOM - IEEE
  Conference on Comput. Commun.}}, pp.~487--495, April 2019.

\bibitem{drcn_gpint}
G.~Simsek, {Ergenç, Doğanalp}, and E.~Onur, ``{Efficient Network Monitoring
  via In-band Telemetry},'' in {\em 17th International Conference on the Design
  of Reliable Communication Networks (DRCN)}, IEEE, 2021.

\bibitem{sqr}
T.~{Qu}, R.~{Joshi}, M.~C. {Chan}, B.~{Leong}, D.~{Guo}, and Z.~{Liu}, ``{SQR:
  In-network Packet Loss Recovery from Link Failures for Highly Reliable
  Datacenter Networks},'' in {\em Proc. of the IEEE 27th International
  Conference on Network Protocols}, pp.~1--12, 2019.

\bibitem{NetView}
Y.~Lin, Y.~Zhou, Z.~Liu, K.~Liu, Y.~Wang, M.~Xu, J.~Bi, Y.~Liu, and J.~Wu,
  ``{NetView: Towards on-demand network-wide telemetry in the data center},''
  {\em Computer Networks}, vol.~180, pp.~1--6, 2020.

\bibitem{sflow}
``{sFlow}.'' \url{{https://sflow.org/}}, 2018.

\bibitem{open_netmon}
N.~L.~M. {van Adrichem}, C.~{Doerr}, and F.~A. {Kuipers}, ``{OpenNetMon:
  Network monitoring in OpenFlow Software-Defined Networks},'' in {\em IEEE
  Network Operations and Management Symposium (NOMS)}, pp.~1--8, 2014.

\bibitem{open_sample}
J.~{Suh}, T.~T. {Kwon}, C.~{Dixon}, W.~{Felter}, and J.~{Carter},
  ``{OpenSample: A Low-Latency, Sampling-Based Measurement Platform for
  Commodity SDN},'' in {\em Proc. of the IEEE 34th Int. Conf. on Distributed
  Computing Systems}, pp.~228--237, 2014.

\bibitem{payless}
S.~R. {Chowdhury}, M.~F. {Bari}, R.~{Ahmed}, and R.~{Boutaba}, ``{PayLess: A
  low cost network monitoring framework for Software Defined Networks},'' in
  {\em {Proc. of the IEEE Network Operations and Management Symposium (NOMS)}},
  pp.~1--9, 2014.

\bibitem{event_triggered_monitoring}
J.~Ku\v{c}era, D.~A. Popescu, H.~Wang, A.~Moore, J.~Ko\v{r}enek, and
  G.~Antichi, ``{Enabling Event-Triggered Data Plane Monitoring},'' in {\em
  {Proc. of the Symposium on SDN Research}}, SOSR '20, pp.~14--26, 2020.

\bibitem{countSketch}
M.~Charikar, K.~Chen, and M.~Farach-Colton, ``{Finding Frequent Items in Data
  Streams},'' vol.~2380, pp.~693--703, 2002.

\bibitem{countMinSketch}
G.~Cormode and S.~Muthukrishnan, ``{An Improved Data Stream Summary: The
  Count-Min Sketch and its Applications},'' {\em J. Algorithms}, vol.~55,
  no.~1, pp.~58--75, 2005.

\bibitem{openSketch}
M.~Yu, L.~Jose, and R.~Miao, ``{Software Defined Traffic Measurement with
  OpenSketch},'' in {\em {Proc. of the 10th {USENIX} Symposium on Networked
  Systems Design and Implementation ({NSDI} 13)}}, pp.~29--42, 2013.

\bibitem{univMon}
Z.~Liu, A.~Manousis, G.~Vorsanger, V.~Sekar, and V.~Braverman, ``{One Sketch to
  Rule Them All: Rethinking Network Flow Monitoring with UnivMon},'' in {\em
  Proc. of the ACM SIGCOMM Conference}, pp.~101--114, 2016.

\bibitem{elasticSketch}
T.~Yang, J.~Jiang, P.~Liu, Q.~Huang, J.~Gong, Y.~Zhou, R.~Miao, X.~Li, and
  S.~Uhlig, ``{Elastic Sketch: Adaptive and Fast Network-Wide Measurements},''
  in {\em Proc. of the ACM SIGCOMM}, {SIGCOMM '18}, pp.~561--575, 2018.

\bibitem{nitroSketch}
Z.~Liu, R.~Ben-Basat, G.~Einziger, Y.~Kassner, V.~Braverman, R.~Friedman, and
  V.~Sekar, ``{NitroSketch: Robust and General Sketch-Based Monitoring in
  Software Switches},'' in {\em Proc. of the ACM SIGCOMM}, pp.~334--350, 2019.

\bibitem{onos-based}
N.~{Van Tu}, J.~{Hyun}, and J.~W. {Hong}, ``{Towards ONOS-based SDN monitoring
  using in-band network telemetry},'' in {\em {Proc. of the 19th Asia-Pacific
  Network Operations and Management Symposium (APNOMS)}}, pp.~76--81, 2017.

\bibitem{int_collector}
N.~V. {Tu}, J.~{Hyun}, G.~Y. {Kim}, J.~{Yoo}, and J.~W. {Hong},
  ``{INTCollector: A High-performance Collector for In-band Network
  Telemetry},'' in {\em Proc. of the 14th International Conference on Network
  and Service Management (CNSM)}, pp.~10--18, 2018.

\bibitem{realtime_intcollector}
J.~Hyun, N.~Van~Tu, J.-H. Yoo, and J.~W.-K. Hong, ``{Real-time and fine-grained
  network monitoring using in-band network telemetry},'' {\em International
  Journal of Network Management}, vol.~29, p.~e2080, 2019.

\bibitem{sINT}
Y.~{Kim}, D.~{Suh}, and S.~{Pack}, ``{Selective In-band Network Telemetry for
  Overhead Reduction},'' in {\em Proc. of the IEEE 7th International Conference
  on Cloud Networking (CloudNet)}, pp.~1--3, 2018.

\bibitem{flex_int}
D.~Suh, S.~Jang, S.~Han, S.~Pack, and X.~Wang, ``{Flexible sampling-based
  in-band network telemetry in programmable data plane},'' {\em ICT Express},
  vol.~6, no.~1, pp.~62--65, 2020.

\bibitem{int_label}
T.~{Pan}, E.~{Song}, C.~{Jia}, W.~{Cao}, T.~{Huang}, and B.~{Liu},
  ``{Lightweight Network-Wide Telemetry Without Explicitly Using Probe
  Packets},'' in {\em Proc. of the IEEE INFOCOM - IEEE Conference on Computer
  Communications Workshops}, pp.~1354--1355, 2020.

\bibitem{pint}
{Ben Basat, Ran and Ramanathan, Sivaramakrishnan and Li, Yuliang and Antichi,
  Gianni and Yu, Minian and Mitzenmacher, Michael}, ``{PINT: Probabilistic
  In-Band Network Telemetry},'' in {\em Proc. of the Annual Conference of the
  ACM Special Interest Group on Data Communication on the Applications,
  Technologies, Architectures, and Protocols for Computer Communication},
  p.~662–680, 2020.

\bibitem{NetVision}
Z.~{Liu}, J.~{Bi}, Y.~{Zhou}, Y.~{Wang}, and Y.~{Lin}, ``{NetVision: Towards
  Network Telemetry as a Service},'' in {\em Proc. of the IEEE 26th
  International Conference on Network Protocols}, pp.~247--248, 2018.

\bibitem{euler_theory}
J.~A. Bondy, {\em {Graph Theory With Applications}}.
\newblock GBR: Elsevier Science Ltd., 1976.

\bibitem{near_optimal_int}
A.~G. Castro, A.~F. Lorenzon, F.~D. Rossi, R.~I. T. d.~C. Filho, F.~M.~V.
  Ramos, C.~E. Rothenberg, and M.~C. Luizelli, ``Near-optimal probing planning
  for in-band network telemetry,'' {\em IEEE Communications Letters}, vol.~25,
  no.~5, pp.~1630--1634, 2021.

\bibitem{int_opt}
D.~{Bhamare}, A.~{Kassler}, J.~{Vestin}, M.~A. {Khoshkholghi}, and J.~{Taheri},
  ``{IntOpt: In-Band Network Telemetry Optimization for NFV Service Chain
  Monitoring},'' in {\em Proc. of the IEEE International Conference on
  Communications}, pp.~1--7, 2019.

\bibitem{marques}
{Jonatas Adilson Marques and Marcelo Caggiani Luizelli and Roberto Iraj{\'a}
  Tavares da Costa Filho and Luciano Paschoal Gaspary}, ``{An
  optimization-based approach for efficient network monitoring using in-band
  network telemetry},'' {\em Journal of Internet Services and Applications},
  vol.~10, no.~1, pp.~1--20, 2019.

\bibitem{kernighan1970efficient}
B.~Kernighan and S.~Lin, ``An efficient heuristic procedure for partitioning
  graphs,'' {\em {The Bell System Technical Journal}}, vol.~49, no.~2,
  pp.~291--307, 1970.

\bibitem{p4Learning}
ETH-Zurich, ``{ETH Zurich P4 Learning Repository}.''
  \url{{https://github.com/nsg-ethz/p4-learning}}, 2020.
\newblock {Accessed: March 2, 2021}.

\bibitem{mininet}
B.~Lantz, B.~Heller, and N.~McKeown, ``{A Network in a Laptop: Rapid
  Prototyping for Software-defined Networks},'' in {\em Proc. of SIGCOMM},
  Hotnets-IX, pp.~19:1--19:6, ACM, 2010.

\end{thebibliography}

%

\end{document}